\documentclass[twocolumn]{aastex631} %

\usepackage{graphicx}	
\usepackage{textcomp}
\usepackage[T1]{fontenc}

\accepted{\today}

\submitjournal{ApJ}

\shorttitle{C--C photocleavage in the ISM}
\shortauthors{Tajuelo-Castilla et al.}

\graphicspath{{./}{figures/}}

\begin{document}

\title{PHOTOCLEAVAGE OF ALIPHATIC C--C BONDS IN THE INTERSTELLAR MEDIUM}

\correspondingauthor{Jos\'{e} \'Angel Mart\'{i}n-Gago; Gonzalo Santoro}
\email{gago@icmm.csic.es, gonzalo.santoro@csic.es}

\author{Guillermo Tajuelo-Castilla}
\affiliation{Instituto de Ciencia de Materiales de Madrid (ICMM), CSIC, c/ Sor Juana Inés de la Cruz 3, E-28049, Madrid, Spain}

\author{Jes\'{u}s I. Mendieta-Moreno}
\affiliation{Departamento de Física Teórica de la Materia Condensada, Facultad de Ciencias, Universidad Autónoma de Madrid, c/Francisco Tomás y Valiente 7, Campus de Excelencia de la Universidad Autónoma de Madrid 28049 Madrid, Spain.}

\author{Mario Accolla}
\affiliation{Osservatorio Astrofisico di Catania, Istituto Nazionale di Astrofisica (INAF), Via Santa Sofia 78, 95123 Catania, Italy.}

\author{Jes\'{u}s M. Sobrado}
\affiliation{Centro de Astrobiología (CAB), CSIC-INTA, Crta. de Torrejón a Ajalvir km 4, E-28850 Torrejón de Ardoz, Madrid, Spain.}

\author{Sofia Canola}
\affiliation{Institute of Physics of the Czech Academy of Sciences, Cukrovarnicka 10, Prague 6, CZ 162 00, Czech Republic.}

\author{Pavel Jel\'{i}nek}
\affiliation{Institute of Physics of the Czech Academy of Sciences, Cukrovarnicka 10, Prague 6, CZ 162 00, Czech Republic.}

\author{Gary J. Ellis}
\affiliation{Instituto de Ciencia y Tecnología de Polímeros (ICTP), CSIC, c/ Juan de la Cierva 3, E-28006, Madrid, Spain.}

\author{Jos\'{e} \'{A}ngel Mart\'{i}n-Gago}
\affiliation{Instituto de Ciencia de Materiales de Madrid (ICMM), CSIC, c/ Sor Juana Inés de la Cruz 3, E-28049, Madrid, Spain.}

\author{Gonzalo Santoro}
\affiliation{Instituto de Estructura de la Materia (IEM), CSIC, c/ Serrano 121, E-28006, Madrid, Spain}

\begin{abstract}

Ultraviolet (UV) processing in the insterstellar medium (ISM) induces the dehydrogenation of hydrocarbons. Aliphatics, including alkanes, are present in different interstellar environments, being prevalently formed in evolved stars; thus, the dehydrogenation by UV photoprocessing of alkanes plays an important role in the chemistry of the ISM,  leading to the formation of unsaturated hydrocarbons and eventually to aromatics, the latter ubiquitously detected in the ISM. Here, through combined experimental results and \textit{ab-initio} calculations, we show that UV absorption (mainly at the Ly-$\alpha$ emission line of hydrogen at 121.6 nm) promotes an alkane to an excited Rydberg state from where it evolves towards fragmentation inducing the formation of olefinic C=C bonds, which are necessary precursors of aromatic hydrocarbons. We show that photochemistry of aliphatics in the ISM does not primarily produce direct hydrogen elimination but preferential C-C photocleavage. Our results provide an efficient synthetic route for the formation of unsaturated aliphatics, including propene and dienes, and suggest that aromatics could be formed in dark clouds by a bottom-up mechanism involving molecular fragments produced by UV photoprocessing of aliphatics.

\end{abstract}


\section{Introduction}\label{sec:intro}
 The molecular inventory of space is comprised of more than 250 identified molecular species \citep{mcguire22} whose formation pathways are very diverse, from gas-phase neutral-neutral reactions in evolved stars to solid-state radiochemistry in molecular clouds \citep{tielens13}. In the interstellar medium (ISM), UV-induced chemistry is particularly relevant and it is considered as responsible for the dehydrogenation of carbonaceous cosmic dust \citep{jenniskens93, jones13, jones17}. In addition, it provides plausible synthetic routes for prebiotic molecules in Dense Molecular Clouds (DMCs), including aminoacids and ribose \citep{bernstein02, munozcaro02, ciesla12, Meinert16, oberg16}, with obvious implications in the emergence of life.

Hydrocarbons are widespread in space \citep{tielens05_book,chiar13,hansen22} and, among them, polycyclic aromatic hydrocarbons (PAHs) account for the capture of up to 20$\%$ of the elemental carbon in the ISM \citep{peeters21}. In particular, the unidentified infrared emission (UIE) bands which fall in the spectral range from 3 to 20 $\mu$m (main bands at 3.3, 6.2, 7.7, 8.6, 11.2 and 12.7  $\mu$m) have generally been assigned to polyaromatic carriers that are small enough to be stochastically heated by the absorption of a single UV photon, which constitutes the polycyclic aromatic hydrocarbon (PAH) hypothesis \citep{leger84, allamandola85, allamandola89, puget89}. The UIEs features are ubiquitously detected in a wide variety of astrophysical regions, including the ISM, star forming galaxies and  extragalatic environments \citep{tielens08,Monfredini2019,Li2020, GarciaBernete2021}. However, the formation mechanism of aromatics is not well constrained and the energetic processing of aliphatic hydrocarbons has been suggested as the driving force for an aliphatic-aromatic transition that leads to the aromatic enrichment of the ISM \citep{goto03, matrajt05,tielens05_ch,tielens13}.

On the other hand, the UIEs are accompanied by IR emission bands at 3.4, 6.85 and 7.25 $\mu$m that are due to aliphatic hydrocarbons \citep{Pinho1995,Yang2013,Jensen2022,Yang2023} and other carriers different from free PAHs have been proposed for the UIEs. These are usually comprised of a mixture of aromatic and aliphatic hydrocarbons and include mixed aromatic and aliphatic organic molecules (MAONs) \citep{kwok11,kwok13} as well as hydrogenated amorphous carbon (HAC) nanoparticles \citep{Duley1988, Jones1990, Jones2012_c, Jones2015, Jones2022}. Indeed, in the diffuse ISM, the 3.4 $\mu$m absorption band along with the weaker absorption features at 6.8 $\mu$m and 7.3 $\mu$m are attributed to the aliphatic component of carbonaceous dust, which is consistent with HAC grains \citep{pendleton02,dartois04}.

Aliphatic hydrocarbons including alkanes are present in different interstellar environments where they are exposed to UV radiation. For instance: long chain aliphatics have been identified in cometary dust \citep{keller06, raponi20}; n-alkanes up to heptane (C$_{7}$H$_{16}$) have been unequivocally detected in-situ by the Rosetta mission in comet 67P/Churyumov-Gerasimenko \citep{schuhmann19}; linear alkanes have also been systematically identified in presolar grains in meteorites \citep{glavin18}; aliphatic hydrocarbons have been recently detected in the samples of the carbonaceous asteroid (162173) Ryugu returned to Earth with a CH$_{2}$/CH$_{3}$ ratio pointing towards longer aliphatic chains than those of meterorites \citep{yabuta23}; to name a few. Nevertheless, it is worth noticing that the presence of alkanes in comets does not necessarily imply their presence in the ISM due to the reprocessing of the interstellar matter in the early solar nebula. Furthermore, C$_{4}$-C$_{6}$ saturated hydrocarbon units are suggested to constitute the aliphatic portion of carbonaceous cosmic dust, weaving the aromatic backbone \citep{pendleton02,dartois05, pino08, kwok11, kwok13}. 

Importantly, aliphatic hydrocarbons, linear alkanes included, are prevalently formed at the conditions of the circumstellar envelopes (CSEs) of carbon-rich evolved stars by the interaction of atomic carbon and H$_{2}$ \citep{martinez20} and these aliphatic molecules are incorporated into the carbonaceous cosmic dust that is expelled towards the interstellar medium. Dehydrogenation of the aliphatic portion of cosmic dust in the ISM increases the C/H ratio in dust grains, which is considered to be a UV-induced process and responsible for the transition from aliphatic-rich to aromatic-rich carbonaceous cosmic dust \citep{pino08, jones13}. 

Spatial mapping of the aliphatic portion of interstellar dust towards the Galactic Centre has found a high variability in the aliphatic content, ranging from 4\% to 25\% of the total carbon abundance depending on the observed source \citep{Godard2012, gunay20}. Thus, aliphatic hydrocarbons can lock as much elemental carbon as PAHs. The observed variability in the aliphatic fraction can be attributed to the different evolutionary stages of the aliphatic-to-aromatic transition \citep{jones17}. Nonetheless, the photon-induced destruction of aliphatics and aliphatic moieties in the ISM is yet to be fully unveiled and the implications of this process on the formation of aromatics is not yet ascertained.

Here, we report that vacuum UV radiation at a photon energy of mainly 10.2 eV (Ly-$\alpha$ emission line of hydrogen) and at low temperatures primarily induces the photocleavage of the C-C bonds in linear alkanes along with subsequent hydrogen transfer between the photofragments; thus, dehydrogenation in the ISM does not occur preferentially through direct hydrogen elimination by the UV radiation field but should substantially proceed as an effective process after aliphatic fragments (carrying hydrogen) are incorporated to the gas phase. In addition, our results can be generalized to several different astrochemical environments, providing a plausible route for the formation of molecular precursors of aromatics in cold environments, where gas-phase chemistry is restricted to barrierless and exoergic reactions. In particular, the mechanism of alkane photo-fragmentation that we present leads to the formation of propene (C$_3$H$_6$) and dienes, whose chemical formation pathways at low temperatures are key for understanding the recent detection of aromatics in dark clouds \citep{mcguire21,cernicharo21}.

\section{Experimental methods and Quantum mechanical calculations }\label{sec:experiments}

All the experiments have been carried out in the INFRA-ICE module \citep{santoro20rev} of the Stardust machine \citep{martinez20,santoro20,accolla21,sobrado23} in ultra-high vacuum (UHV) conditions (base pressure at room temperature: 3 $\times$ 10$^{-10}$ mbar). 

We performed two different independent irradiation experiments. The first one consisted in the deposition of linear hexane (C$_6$H$_{14}$; Sigma-Aldrich; purity > 99\%) and subsequent UV-irradiation. The second was intended to generalize the results and consisted in the deposition and subsequent UV-irradiation of linear undecane (C$_{11}$H$_{24}$; Sigma-Aldrich; purity > 99\%). In both cases, alkane vapours were deposited on infrared transparent KBr substrates at 14 K. Prior to introducing vapours in the chamber, alkanes were further purified by three pump-thaw cycles. The column density (number of molecules per cm$^2$), \textit{N}, of the deposited alkanes was calculated from the IR spectrum using the CH$_2$/CH$_3$ stretching modes region (2800-3000 cm$^{-1}$) according to

\begin{equation} \label{eq1}
    N=\int\frac{\tau(\nu)d\nu}{A}
\end{equation}

\noindent
where $\tau$ is the optical depth and \textit{A} the band strength of the overall CH$_2$/CH$_3$ stretching modes. We used \textit{A} values of 7.2 $\times$ 10$^{-17}$ cm molecule$^{-1}$ for C$_6$H$_{14}$ \citep{matrajt05} and 1.3 $\times$ 10$^{-16}$ cm molecule$^{-1}$ for C$_{11}$H$_{24}$ \citep{dartois04} which lead to column densities of (4.6 $\pm$ 0.9) $\times$ 10$^{16}$ molecules cm$^{-2}$ and (4.3 $\pm$ 0.9) $\times$ 10$^{16}$ molecules cm$^{-2}$, respectively, corresponding to about 45 monolayers (1 ML $\approx$  10$^{15}$ molecules cm$^{-2}$).

After deposition, solid C$_6$H$_{14}$ and C$_{11}$H$_{24}$ were irradiated by UV photons using a H$_2$-flowing discharge lamp (UVS 40A2, Prevac) operating at 60 W. At the selected working conditions, the spectrum of the lamp corresponds predominantly to the Lyman-$\alpha$ line of atomic hydrogen at 121.6 nm (10.2 eV) with contributions from the emission of molecular hydrogen at around 160 nm (7.8 eV). Hydrogen discharge lamps favouring Lyman-$\alpha$ emission have been shown to satisfactorily simulate the UV field of the ISM \citep{jenniskens93} and have also been used to simulate the secondary UV field in DMCs \citep{alata2014} and Photon-Dominated Regions (PDRs) \citep{alata15}. As our lamp is windowless contributions from Lyman-$\beta$ and Lyman-$\gamma$ lines at 102.6 nm (12.1 eV) and 97.3 nm (12.7 eV) are also present. Therefore, windowless UV discharge lamps, more closely reproduce the ISM UV field as they cover the UV emission in the 91.2–115 nm range. Windowed lamps usually employ MgF$_2$ windows which shows a cut-off at wavelengths below 115 nm \citep{chen13}. 

At the selected working conditions, the photon flux integrated over the whole spectral range is 6.2 $\times$ 10$^{14}$ ph s$^{-1}$ cm$^{-2}$ \citep{santoro20rev}. Total UV fluences of ca. 10$^{19}$ ph cm$^{-2}$ were employed. Considering a photon field in the diffuse ISM of $\sim$ 8 $\times$ 10$^7$ ph s$^{-1}$ cm$^{-2}$ \citep{mathis83}, the employed fluence corresponds to $\sim$ 10$^3$ - 10$^4$ years in the diffuse ISM. In the case of DMCs, the secondary UV field is estimated as 10$^4$ ph s$^{-1}$ cm$^{-2}$ \citep{CecchiPestellini92}; thus, the total UV fluence used in the experiments corresponds to $\sim$ 3 $\times$ 10$^7$ years, a time similar to the lifetime of molecular clouds \citep{chevance19}. Nevertheless, it should be noted that the photon flux in our experiments is orders of magnitude higher than that in the ISM and DMCs, what might play a role as, e.g., relaxation between photon absorption events can be impeded.

During the complete UV irradiation, transmission IR spectra were concurrently acquired each 200 s using a vacuum VERTEX 70V spectrometer (Bruker) with a liquid-nitrogen cooled mercury-cadmium-telluride (MCT) detector. The complete optical path is kept under vacuum (10$^{-1}$ mbar). The spectral resolution was set to 2 cm$^{-1}$ and 128 scans were co-added for each spectrum. The ZnSe windows that are used to isolate the vacuum of the spectrometer to the UHV of the sample chamber strongly decrease the sensitivity in the spectral range below 850 cm$^{-1}$.

From the IR spectra we have calculated the effective cross-sections of photodestruction, $\sigma_{des}$, and photoformation, $\sigma_{form}$, for several molecular moieties. The effective cross-sections are derived considering first-order reaction kinetics according to the following expressions \citep{cottin03, loeffler05,MartinDomenech15}:

\begin{equation}\label{eq2}
    \tau(t)=\tau_{ss}+(\tau_0 - \tau_{ss}) e^{-\sigma_{des}\phi t}
\end{equation}

\begin{equation}\label{eq3}
    \tau(t)=\tau_{ss}(1-e^{-\sigma_{form}\phi t})
\end{equation}

\noindent
where $\tau$ denotes the integrated optical depth of the selected IR band, $\tau_{ss}$ the steady state optical depth, $\tau_0$ the initial integrated optical depth, $\phi$ the photon flux and $t$ the irradiation time. Effective cross-sections encompass all the possible formation/destruction pathways and therefore do not distinguish among  different chemical routes. 

Thermal Programmed Desorption (TPD) measurements for C$_6$H$_{14}$ were performed after the UV irradiation at a heating rate of 1 K min$^{-1}$ using a Lakeshore 335 temperature controller. A PrismaPlus QMG 220 M2 (Pfeiffer) mass spectrometer continuously monitored the desorbed gaseous species from m/z = 1 to m/z = 200 and a complete mass spectrum was acquired every 1.2 K. From the mass spectra, TPD curves were derived at selected m/z values. TPD measurements of an identical sample without UV exposure was also acquired for comparison purposes.

Quantum mechanical calculations were performed at different levels of theory. To investigate the softening of the C-C bonds in C$_6$H$_{14}$, we fixed the C2-C3 bond length and relaxed the remaining degrees of freedom. This was performed for the neutral ground state (S$_0$; C$_6$H$_{14}$), the cation ground state (S$_0$; C$_6$H$_{14}$$^{+}$) and neutral first excited state (S$_1$; C$_6$H$_{14}$). 

Energy barriers for neutral and cation states have been calculated using Density Functional Theory (DFT) \citep{lewis11} using the BLYP exchange-correlation functional \citep{lee88} with D3 corrections \citep{grimme11} and norm conserving pseudopotentials. We employed a basis set of optimized numerical atomic-like orbitals (NAOs) \citep{basanta07} with a 1s orbital for H and sp$^3$ orbitals for C atoms. The energy for the barriers were calculated by fixing the reactions coordinates and relaxing the geometries of the molecules. 

For the analysis of the excited states DFT time-dependent DFT (TD-DFT) quantum-chemical calculations were performed with Gaussian16 software \citep{g16} employing CAM-B3LYP functional and 6-31++G* basis set. The neutral and cation structures of C$_6$H$_{14}$ have been relaxed in their ground state and the first hundred vertical excited states of the neutral molecule have been calculated. Full relaxation of the neutral alkane on the first excited state surface was performed. For neutral ground state (S$_0$; C$_6$H$_{14}$), cation ground state (S$_0$; C$_6$H$_{14}$$^+$) and neutral first excited state (S$_1$; C$_6$H$_{14}$), the energy scan was performed constraining the C2-C3 bond length and relaxing all the other degrees of freedom. The calculations for the energy scan have also been performed using Gaussian16 software \citep{g16}. Finally, the ionization potential is estimated as the difference between the energy at the ground state neutral and that of the cation ground state, at a fixed geometry of the neutral state.

\section{Results}\label{sec:results}
\subsection{Formation of new chemical species during UV irradiation of linear alkanes}\label{sec:IR}

\begin{figure*}[hbt!]
    \centering
    \includegraphics[width=0.75\textwidth]{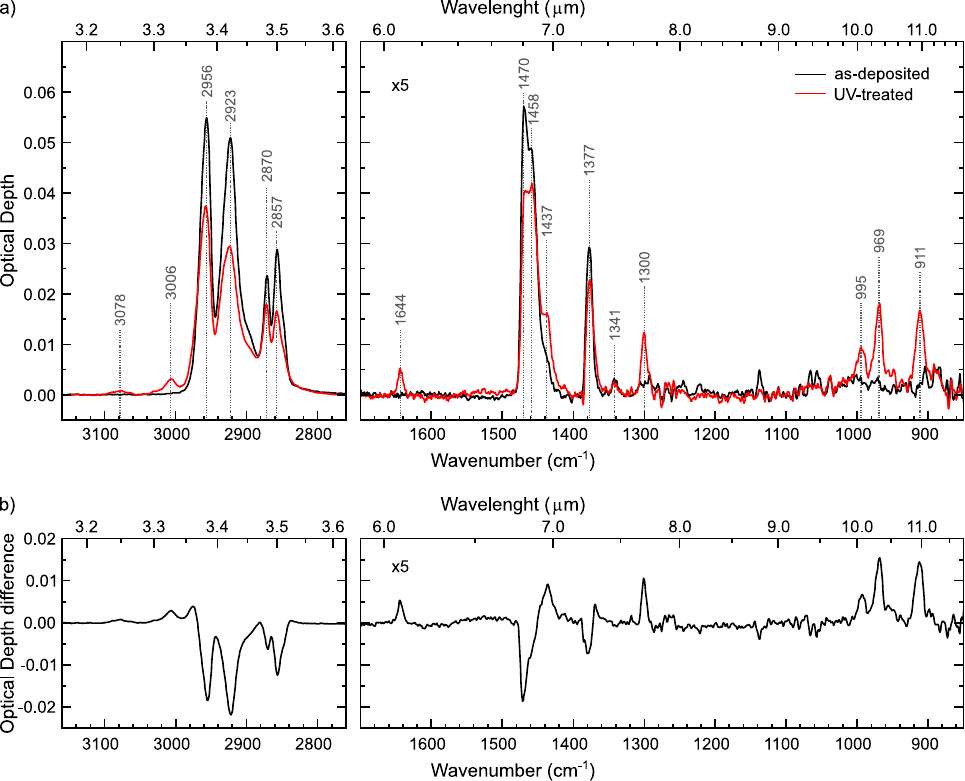}
    \caption{a) IR spectra of C$_6$H$_{14}$ at 14 K both as-deposited and after a UV fluence of 9.8 $\times$ 10$^{18}$ ph cm$^{-2}$. The position of the most prominent absorption bands is indicated in the figure. b) Difference spectrum between the UV-treated and the as-deposited spectra.}
    \label{fig:IR}
\end{figure*}

To investigate the UV photochemistry of linear alkanes at low temperature and at the conditions of the ISM we irradiated both linear hexane (C$_6$H$_{14}$) and undecane (C$_{11}$H$_{24}$) mainly with the Lyman-$\alpha$ emission of hydrogen at 10.2 eV ($\lambda$=121.6 nm). Figure~\ref{fig:IR}a shows the IR spectra of solid amorphous C$_6$H$_{14}$ both as-deposited and after irradiation with a UV fluence of ca. 10$^{19}$ ph cm$^{-2}$. To highlight the changes in the spectra upon irradiation, Figure~\ref{fig:IR}b shows the difference spectrum. 

A clear reduction in the bands associated with C$_6$H$_{14}$ is observed (see Appendix~\ref{app_assign} for IR band assignment) along with the emergence of new absorption features that reveal the formation of olefinic moieties, both of vinyl (-CH=CH$_2$) and \textit{trans}-vinylene (-CH=CH-) character. In particular, the bands at 1644 cm$^{-1}$ and 3078 cm$^{-1}$ are ascribed to the C=C and CH stretching modes of olefins whereas the doublet at 995 cm$^{-1}$ and 911 cm$^{-1}$ and the band at 969 cm$^{-1}$ are very characteristic of vinyl (-CH=CH$_2$) and \textit{trans}-vinylene (-CH=CH-) moieties, respectively \citep{socrates}. The band at 1437 cm$^{-1}$, which is observed to increase upon UV irradiation, can be attributed to methylene (CH$_2$) deformation in the presence of adjacent unsaturated groups \citep{socrates}. This assignment becomes clearer when considering the irradiation of crystalline C$_6$H$_{14}$ (see Appendix ~\ref{app:cristC6H14}).

The IR spectra also shows the formation of methane (CH$_4$) as revealed by the IR bands at 1300 cm$^{-1}$ and 3006 cm$^{-1}$ \citep{gerakines96}, which might imply the formation of CH$_3$ radicals upon UV exposure. However, our \textit{ab initio} calculations show that CH$_4$ can be directly formed as consequence of C-C photocleavage (see Section~\ref{sec:calc} and Appendix~\ref{app:barriers}) indicating that the photochemistry of C$_6$H$_{14}$ may not be mediated by radical species.

A list of the new absorption features after UV irradiation along with its assignment is given in Table~\ref{table:ir}. Identical results were obtained for C$_{11}$H$_{24}$ (see Appendix~\ref{app:C11H24}) implying that the mechanism for olefin formation is not restricted to C$_6$H$_{14}$ but general to mid- and long-chain linear alkanes.
\begin{deluxetable}{ccl}[hbt!]
    \label{table:ir}
    \tablecaption{New IR bands upon UV irradiation of C$_6$H$_{14}$ along with band assignment.}
    \tablewidth{\columnwidth}
    \tablehead{
    \colhead{Wavenumber} & \colhead{Wavelenght} & \colhead{Assignment$^{(a)}$} \\
    \colhead{cm$^{-1}$} & \colhead{$\mu$m}}

        \startdata
        3078 & 3.25  & $\nu_{as}$ CH (=CH)            \\
        3006 & 3.33  & $\nu_{as}$ CH (CH$_4$)            \\
        1644 & 6.08  & $\nu$ CH=CH                 \\
        1437 & 6.96  & $\delta$ CH$_2$$^{(b)}$               \\
        1300 & 7.69  & $\delta$ CH (CH$_4$)              \\
        995  & 10.04 & $\gamma_{oop}$ CH (-CH=CH$_2$) \textit{vinyl} \\
        969  & 10.25 & $\gamma_{oop}$ CH (-CH=CH-) \textit{trans} \\
        911  & 10.98 & $\gamma_{oop}$ CH (-CH=CH$_2$) \textit{vinyl} \\
        \enddata
    \tablecomments{The vibrational modes are abbreviated as follows: $\nu$: stretching; $\delta$: deformation; $\gamma$: wagging; as: asymmetric; oop: out-of-plane; $^{(a)}$ Assignments from \citep{socrates,gerakines96}; $^{(b)}$ This band is attributed to the deformation of CH$_2$ in the presence of adjacent unsaturated groups.}
\end{deluxetable}

\begin{figure}[hbt!]
    \centering
    \includegraphics[width=\columnwidth]{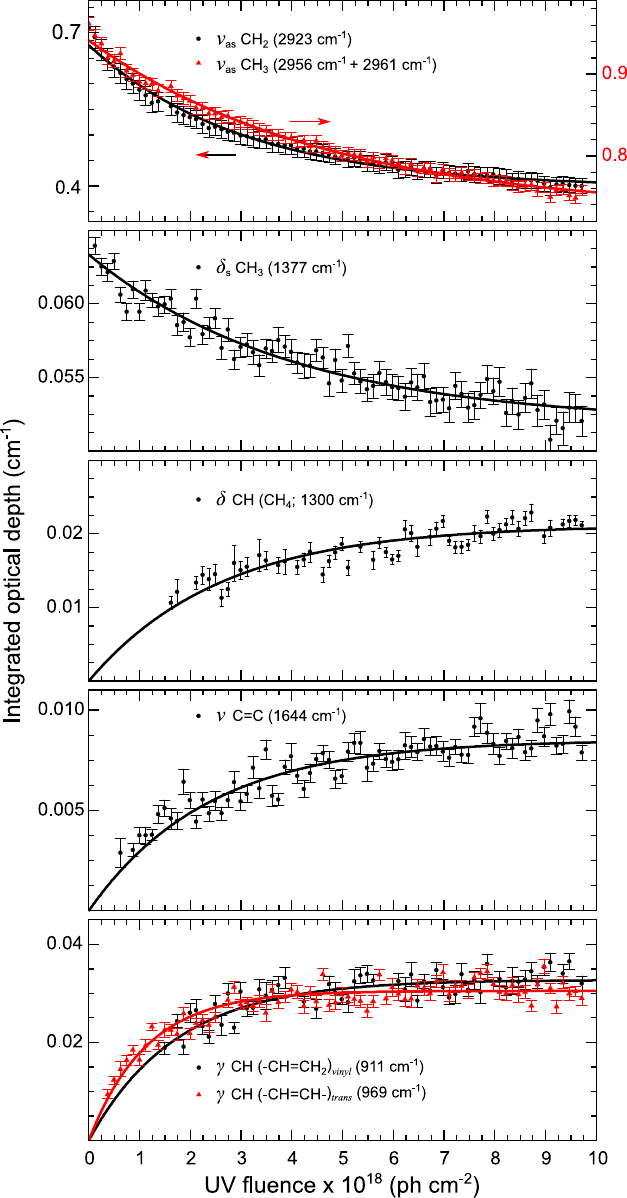}
    \caption{Evolution of the integrated optical depth of selected IR absorption features with UV fluence. Solid lines correspond to the fitting of the experimental data to Eqs.~\ref{eq2} and ~\ref{eq3}}
    \label{fig:cross_sections}
\end{figure}

From the evolution of the IR absorption features with UV fluence (Fig.~\ref{fig:cross_sections}), we have derived the effective destruction cross-sections, $\sigma_{des}$, of CH$_2$ and CH$_3$ aliphatic moieties for C$_6$H$_{14}$, which show values of 3.2 $\times$ 10$^{-19}$ cm$^2$ ph$^{-1}$ and 2.1-2.5 $\times$ 10$^{-19}$ cm$^2$ ph$^{-1}$, respectively. The higher value observed for CH$_2$ destruction indicates that the cleavage of C-C bonds is more likely to occur in the molecule backbone, a result that is further confirmed by the \textit{ab initio} calculations (see Section~\ref{sec:calc}).

We have also derived the effective formation cross section, $\sigma_{form}$, of CH$_4$ (3.8 $\times$ 10$^{-19}$ cm$^2$ ph$^{-1}$) and olefinic moieties. Interestingly, the formation cross-section of \textit{trans}-vinylene moieties (-CH=CH-) (8.8 $\times$ 10$^{-19}$ cm$^2$ ph$^{-1}$) is higher than that of vinyl moieties (-CH=CH$_2$) (5.8 10$^{-19}$ cm$^2$ ph$^{-1}$), which might suggest the preferential formation of C=C at the molecule backbone. However, this result should be taken with caution since the absorption coefficients for the IR bands analysed will depend on the particular unsaturated hydrocarbon and on its chemical environment, which makes it difficult to obtain definite results. The obtained values for the cross-sections are listed in Table ~\ref{table:cross_sections}. We note that due to the absence of isolated IR bands solely ascribed to C$_6$H$_{14}$, the photolysis rate of hexane cannot be derived from the IR spectra. Only the overall effective decrease in CH$_2$/CH$_3$ saturated aliphatic moieties can be obtained, i.e., the reduction in the number of CH$_2$/CH$_3$ moieties as a result of the formation of olefins.
\begin{table}[hbt!]
    \caption{Photodestruction and photoproduction effective cross-sections of C$_6$H$_{14}$}              
    \tablewidth{\columnwidth}
    \label{table:cross_sections}      
    \centering
    \begin{tabular} {c c c}
        
        \hline\hline                        
        Vibrational mode$^{(a)}$ & Wavenumber & $\sigma_{des}$$^{(b)}$ \\
        & cm$^{-1}$ &  cm$^2$ ph$^{-1}$ $\times$ 10$^{-19}$\\
        \hline
        $\nu_{as}$ CH$_3$ & 2956 + 2961 & 2.1 \\
        $\nu_{as}$ CH$_2$ & 2923          & 3.2 \\
        $\delta_{s}$ CH$_3$  & 1377          & 2.5 \\
        \hline                        
        Vibrational mode$^{(a)}$ & Wavenumber & $\sigma_{form}$$^{(b)}$ \\
        & cm$^{-1}$ &  cm$^2$ ph$^{-1}$ $\times$ 10$^{-19}$\\
        \hline
        $\nu$ C=C                     & 1644 & 4.2   $^{(c)}$ \\
        $\delta$ CH (CH$_4$)                & 1300 & 3.8       \\
        $\gamma_{oop}$   CH (-CH=CH-) \textit{trans} & 969  & 8.8       \\
        $\gamma_{oop}$   CH (-CH=CH$_2$) \textit{vinyl} & 911  & 5.8    \\
        \hline
    \end{tabular}
    \tablecomments{$^{(a)}$ For the abbreviation of the vibrational modes the reader is referred to Table ~\ref{table:ir}. s: symmetric}  $^{(b)}$ The error of the cross sections is estimated at 20$\%$.$^{(c)}$The main contribution to this value comes from vinyl moieties since the C=C stretching mode of \textit{trans} alkene moieties is weak if not absent.
    
\end{table}

\subsection{Thermal desorption after UV irradiation}\label{sec:TPD}

\begin{figure}[hbt!]
    \centering
    \includegraphics[width=\columnwidth]{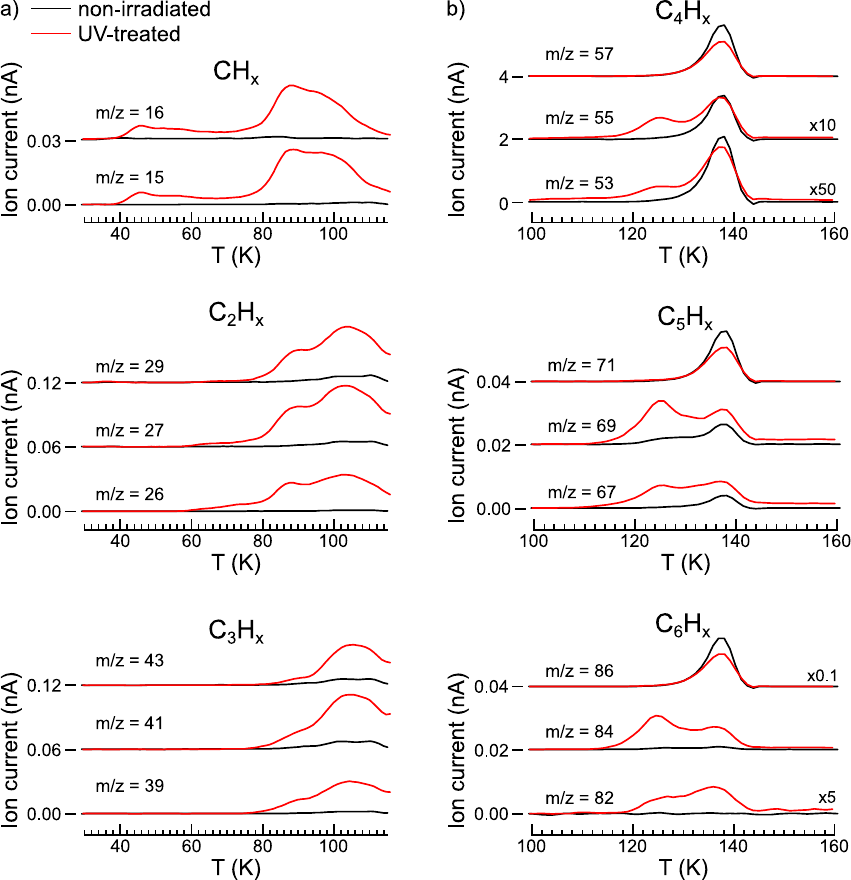}
    \caption{TPD results after the UV irradiation of C$_6$H$_{14}$ with a fluence of 9.8 $\times$ 10$^{18}$ ph cm$^{-2}$ in the temperature ranges a) 30-120 K and b) 100-160 K. The results from non-irradiated C$_6$H$_{14}$ are also provided for comparison.}
    \label{fig:TPD}
\end{figure}

The formation of olefin moieties is further confirmed by Thermal Programmed Desorption (TPD) measurements. The results for some selected m/z values characteristic of aliphatic C$_n$H$_m$ (1 $\leq$ n $\leq$ 6; m $\leq$ 14) molecular species are shown in Figure~\ref{fig:TPD}, where the TPD measurements of non-irradiated C$_6$H$_{14}$ are also shown for comparison purposes. The desorption of CH$_4$ at around 45 K is evident by the increase in m/z =15 and 16 (Fig.~\ref{fig:TPD}a). As the temperature increases C$_2$H$_x$ and C$_3$H$_x$ species desorb with maximum desorption temperatures of 89 K and 105 K, respectively. We also detected C$_2$H$_x$ and C$_3$H$_x$ unsaturated species by their characteristic signals at m/z = 26, 27 and m/z = 39, 41. We note that m/z = 41 corresponds to the most intense signal from C$_3$H$_6$.

At higher temperatures, clear signatures of single and double C=C bonds in C$_4$H$_x$, C$_5$H$_x$ and C$_6$H$_x$ species are observed. The desorption temperature of olefins occurs at lower temperatures regarding their fully saturated counterparts and exhibit electron-impact dissociation patterns with mass peaks at $\Delta$(m/z) = -2 with respect to the corresponding alkane \citep{gross17}. The peaks at m/z = 53, 55 (related to C$_4$H$_6$ and C$_4$H$_8$, respectively), m/z = 67, 69 (related to C$_5$H$_8$ and C$_5$H$_{10}$, respectively) and at m/z = 82, 84 (C$_6$H$_{10}$ and C$_6$H$_{12}$) shows maxima at a temperature of 126 K (Fig.~\ref{fig:TPD}b), once again verifying C=C bond formation. The peaks at m/z = 82 and at m/z = 84 are unambiguously ascribed to the parent molecules C$_6$H$_{10}$ and C$_6$H$_{12}$. The former confirms that dienes are formed.

\begin{figure}[hbt!]
    \centering
    \includegraphics[width=\columnwidth]{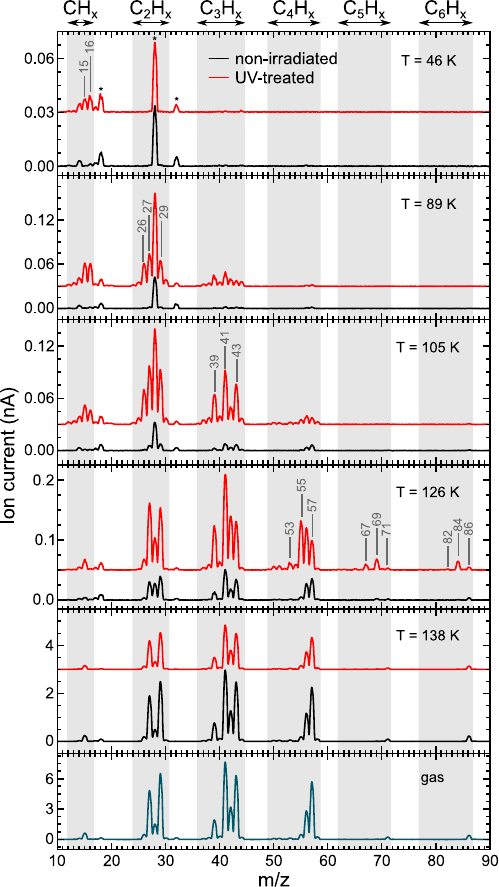}
    \caption{Mass spectra during the desorption of non-irradiated and UV-treated solid C$_6$H$_{14}$. The desorption temperature of each spectrum is indicated. The gas phase C$_6$H$_{14}$ spectrum is shown for comparison purposes. Peaks labelled with * at m/z = 18, 28 and 32 corresponds to residual H$_2$O, CO and O$_2$ gases in the chamber, respectively.}
    \label{fig:Fig_mass_spectra}
\end{figure}

Fig.~\ref{fig:Fig_mass_spectra} shows a comparison of the mass spectra of non-irradiated and UV-treated C$_6$H$_{14}$ at selected desorption temperatures, corresponding to the most prominent desorption temperatures of C$_n$H$_m$ (1 $\leq$ n $\leq$ 6) hydrocarbons. The gas phase spectrum of C$_6$H$_{14}$ is also provided for comparison purposes. We note that we did not observe desorbed species at m/z > 86, which indicates that polymerization from C$_6$H$_{14}$ fragments does not occur or at least is only a residual process. However, as our results are related to irradiated, isolated alkanes caution has to be taken to directly extrapolate them to the post-irradiation behaviour of alkane moieties in hydrogenated amorphous carbon grains. 

\subsection{Mechanism of C-C photocleavage}\label{sec:calc}

In order to gain insight into the C-C photocleavage mechanism we performed quantum-mechanical calculations. The calculated ionization energy of C$_6$H$_{14}$ is 10.8 eV, in agreement with experimental results for the gas-phase ionization of C$_6$H$_{14}$, for which ionization energies of 10.1-10.6 eV have been reported \citep{hoogerbrugge89,steenvoorden91}. Therefore, the ionization potential lies on the same energy range of the main photon energy used in the experiments (10.2 eV). To account for the possible excited states of C$_6$H$_{14}$ upon UV excitation we have simulated the absorption spectrum of C$_6$H$_{14}$ using TD-DFT calculations (Appendix~\ref{app:excited_states}). These show spectroscopically active excited states with an energy onset of ca. 8 eV \citep{morisawa12,mao19} and intense absorption between 9.5-10.5 eV, where the main experimental photon energy lies.

In the case of alkanes, the low-lying excited states present a Rydberg character, displaying an electronic distribution far from the nuclei with a loosely bound excited electron \citep{morisawa12,mao19}. We verified this on C$_6$H$_{14}$ by considering the lowest excited state (S$_1$) which is represented by the transition from the Highest Occupied Molecular Orbital (HOMO) to the Lowest Unoccupied Molecular Orbital (LUMO) (see Appendix~\ref{app:excited_states}). S$_1$ presents an extended electron spatial distribution far from the nuclei due to the promotion of an electron from the HOMO to the LUMO orbital, having a Rydberg character. Because of this particular electronic structure, the electron can be considered as effectively detached and some characteristics of excited Rydberg states converge to those of the related cation, in particular when describing the reactivity of the excited states \citep{lipsky81}. To verify this assumption, we compared the computed relaxed geometries of neutral C$_6$H$_{14}$ in the ground state S$_0$, cation C$_6$H$_{14}$$^+$ in the ground state S$_0$ and neutral C$_6$H$_{14}$ in the S$_1$ excited state (for simplicity, in the following we will refer to them as S$_0$-neutral, S$_0$-cation and S$_1$-neutral, respectively).

\begin{figure*}[hbt!]
    \centering
    \includegraphics[width=0.75\textwidth]{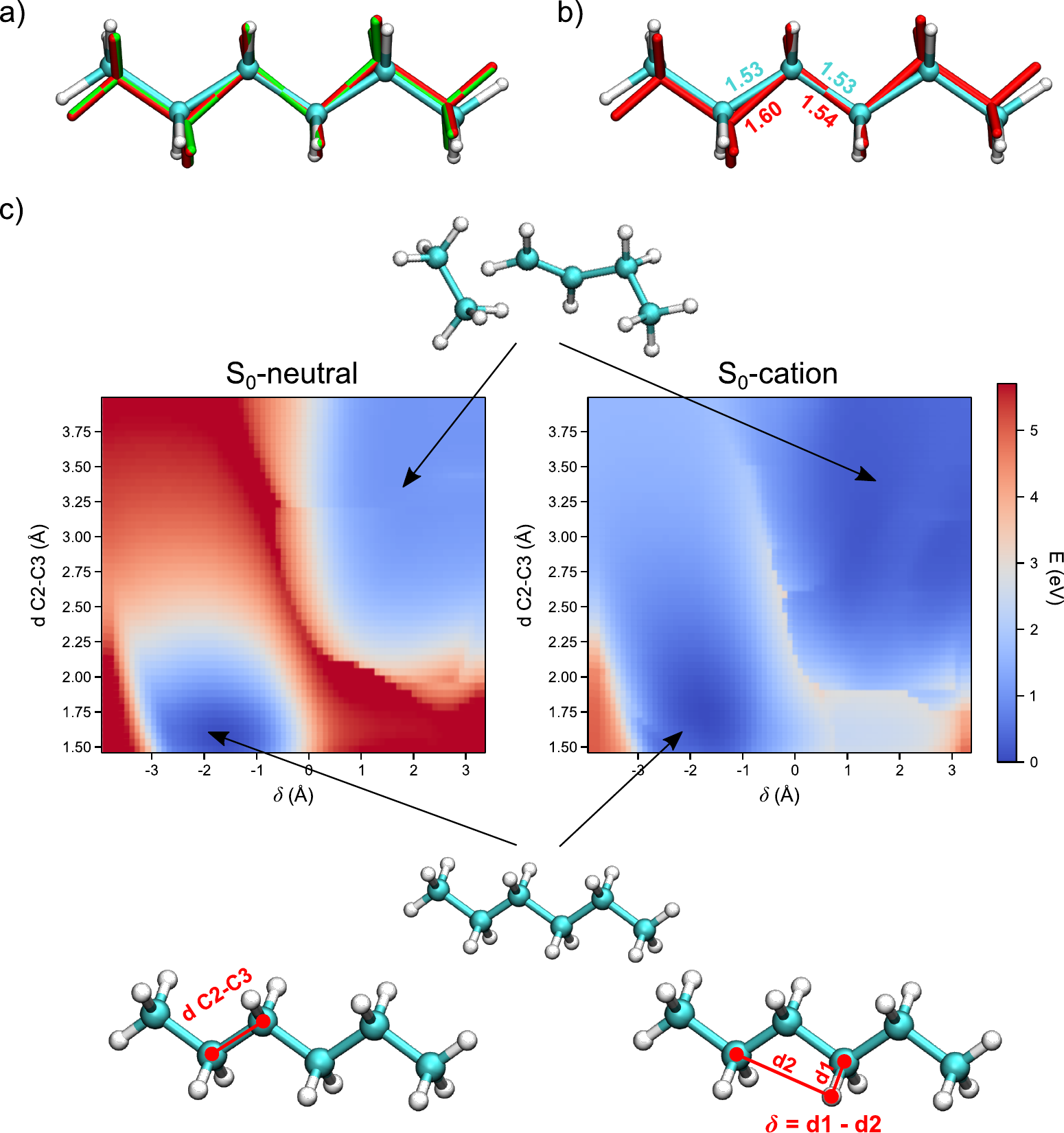}
        \caption{a) Comparison of the relaxed geometries of C$_6$H$_{14}$ in the neutral ground state S$_0$ (ball and stick), the first excited state S$_1$ (green) and the cation C$_6$H$_{14}$$^+$ in the ground state S$_0$ (red). b) Comparison of the relaxed geometries of ground state S$_0$ of C$_6$H$_{14}$ (ball and stick) and ground state S$_0$ of C$_6$H$_{14}$$^+$ (red) highlighting the elongation of the C2-C3 and C3-C4 bonds. The distances are given in Å. c) Energy landscapes for the reaction C$_6$H$_{14}$ $\rightarrow$ C$_2$H$_6$ + C$_4$H$_8$ for neutral C$_6$H$_{14}$ and cation C$_6$H$_{14}$$^+$ molecules, both in the ground state S$_0$. The reaction coordinates are depicted in the bottom of the figure.}
    \label{fig:theory_main}
\end{figure*}

Interestingly, the relaxed geometry of S$_1$-neutral shows marked similarities with that of S$_0$-cation (Fig.~\ref{fig:theory_main}a) and when comparing the relaxed geometries of S$_0$-neutral to those of S$_0$-cation and S$_1$-neutral, we found that all C-C bonds lengthen. This indicates a softening of all C-C bonds (Table~\ref{table:softening}). In particular, the C2-C3 bond has a remarkable elongation: 1.60 Å for S$_0$-cation/S$_1$-neutral vs. 1.53 Å for S$_0$-neutral.
\begin{table}[hbt!]
\caption{C-C bond distances of C$_6$H$_{14}$ in the neutral S$_0$, cation S$_0$ and neutral S$_1$ states.}          
    \label{table:softening}      
    \centering
\begin{tabular}{r c c c}
 \hline\hline 
      & \multicolumn{3}{c}{Bond length  (Å)} \\
      \hline
      & S$_0$-neutral & S$_0$-cation & S$_1$-neutral \\
      \hline
C1-C2 & 1.527      & 1.535     & 1.533      \\
C2-C3 & 1.528      & 1.605     & 1.600      \\
C3-C4 & 1.528      & 1.540     & 1.540 \\
\hline
\end{tabular}
\end{table}

\begin{figure}[hbt!]
    \centering
    \includegraphics[width=\columnwidth]{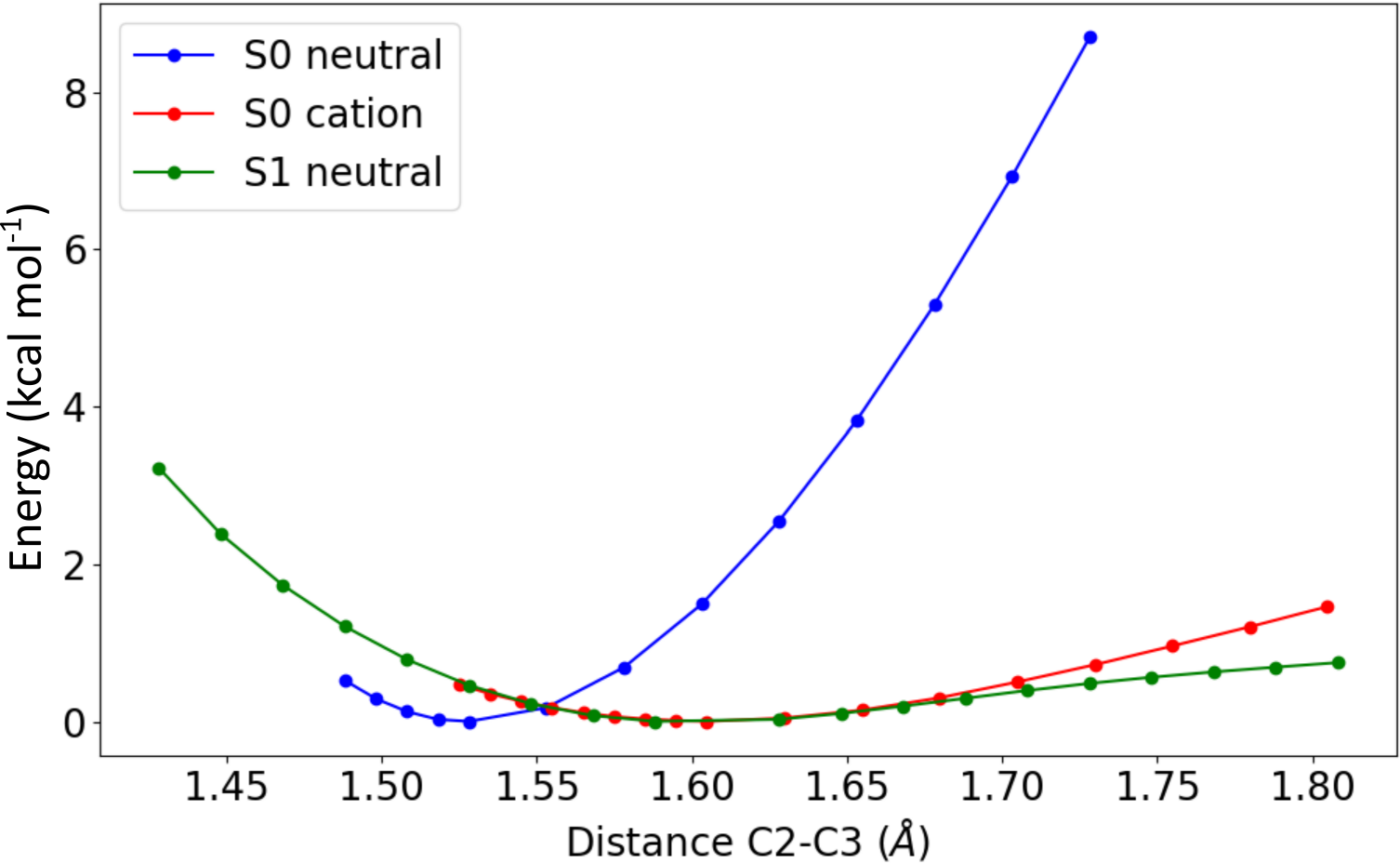}
    \caption{Energy variation upon scan along the C2-C3 bond length: the relaxed electronic structure of C$_6$H$_{14}$ in ground neutral state (S$_0$-neutral, blue), cation in the ground state (S$_0$-cation, red) and first neutral excited state (S$_1$-neutral, green) (CAM-B3LYP/6-31++G*).}
    \label{fig:softening}
\end{figure}

We also calculated the energy required to elongate the C2-C3 bond. We choose this bond as is the one exhibiting the higher elongation with respect to the neutral ground state but similar results are expected for the other C-C bonds. We found that the elongation energy is significantly lower for both S$_0$-cation and S$_1$-neutral with respect to S$_0$-neutral, with that for S$_1$-neutral even further reduced (see Fig.~\ref{fig:softening}). From these calculations it is evident that both S$_0$-cation and S$_1$-neutral facilitate the C-C cleavage: the energy gain needed to elongate the C2-C3 distance is considerably lower than in the case of the neutral ground state. Both S$_0$-cation and S$_1$-neutral states behave similar at regions close to the energy minima, in accordance with the very similar relaxed geometries of both systems (Fig.~\ref{fig:theory_main}a), showing that the softening of the C2-C3 bond upon excitation (S$_1$-neutral) or removal of an electron from the HOMO valence orbital (S$_0$-cation) is similar.

Overall, the shown similarities in terms of structure and electronic properties between S$_1$-neutral and S$_0$-cation, indicate that the behaviour of the excited Rydberg states can be satisfactorily approximated by that of the cation and this is the approach adopted here. Thus, to model the fragmentation reactions occurring upon photoexcitation, the S$_1$-neutral Rydberg state has been approximated by the S$_0$-cation state.

Using this approach, we have calculated the reaction energy landscapes for C-C cleavage reactions of C$_6$H$_{14}$ both in the S$_0$-neutral and S$_0$-cation states as well as for C-H breaking reactions leading to the elimination of H$_2$ (see Appendix~\ref{app:barriers}). In Figure~\ref{fig:theory_main}c we show the results for the C2-C3 rupture and subsequent hydrogen relocation leading to the formation of C$_2$H$_6$ and C$_4$H$_8$.

Both in the S$_0$-neutral and S$_0$-cation states there is a considerable energy barrier for the reaction but this is drastically reduced from 5.00 eV to 1.80 eV from the neutral to the cationic state. Thus, fragmentation in the S$_0$-cation state becomes easier than in the S$_0$-neutral state. Here, we recall that we are approximating the excited S$_1$-neutral Rydberg state by the S$_0$-cation state; thus, our results show that the fragmentation of C$_6$H$_{14}$ is favoured when proceeding through electronically excited states. According to the energy landscape, the reaction for the formation of C$_2$H$_6$ and C$_4$H$_8$ in the S$_0$-cation state proceeds by the elongation of the C2-C3 bond and the subsequent relocation of an H atom from C4 to C2, which induces the formation of a C=C bond between atoms C3 and C4, in agreement with the formation of vinyl groups observed by IR spectroscopy.

We have found a marked reduction in the energy barriers (by factors higher than 2) for all the calculated fragmentation reactions when considering S$_0$-cation states with respect to the same reactions through S$_0$-neutral states using BLYP-D3/NAO with Fireball DFT (see Appendix~\ref{app:barriers}). The reaction barriers are listed in Table~\ref{table:barriers}. 

Importantly, in the S$_0$-cation states all the barrier energies for C-C bond breaking are lower than that for H$_2$ elimination and the reduction in energy barriers is more pronounced (both in absolute and relative values) for all the reactions involving C-C cleavage.These results might reflect the lower bond energies of aliphatic C-C bonds ($\sim$ 3.8 eV) regarding that of aliphatic C-H bonds ($\sim$ 4.3 eV) \citep{Duley2000, jones12} and prove that C-C cleavage is preferential over C-H bond breaking upon electronic excitation, in line with the experimental observation. In addition, the simulation of the energy landscape shown in Figure~\ref{fig:app_barriers_H2} (Appendix ~\ref{app:barriers}) demonstrates that the radical formation through C$_6$H$_{14}$ $\rightarrow$ C$_6$H$_{13}$ + H (energy barrier of $\sim$ 4.5 eV) is less favourable than C-C bond cleavage from an excited Rydberg state.

\begin{table*}[hbt!]
\caption{Energy barriers for the fragmentation of C$_6$H$_{14}$ for the formation of different products.}              
    \label{table:barriers}      
    \centering
    \begin{tabular}{rccccc}
    \hline\hline
               & \multicolumn{5}{c}{Reaction energy barriers (eV)}                   \\
    \hline
    Products   & CH$_4$ + C$_5$H$_{10}$ & C$_2$H$_4$ + C$_4$H$_{10}$ & C$_2$H$_6$ + C$_4$H$_8$ & C$_3$H$_6$ + C$_3$H$_8$ & C$_6$H$_{12}$ + H$_2$ \\
    \hline
    S$_0$-neutral & 5.64        & 5.42         & 5.00        & 5.85        & 5.85       \\
    S$_0$-cation  & 2.38        & 1.95         & 1.80        & 2.65        & 2.80      \\
    \hline
\end{tabular}
\end{table*}

Despite the aforementioned arguments towards a preferential C-C photocleavage and although the energy barriers are considerably reduced when the reactions proceed through excited electronic states, the energy barriers are still high. Our calculations show that the geometrical relaxation from the S$_0$-cation state provides 0.5 eV of thermal energy, which is still insufficient for the fragmentation of the molecule. Nevertheless, the dissociation barrier can be surpassed by the energy provided through the internal conversion from high energy excited states to low-lying Rydberg states. Part of the difference in energy from an excited state at the main photon experimental excitation (10.2 eV) to the S$_1$-neutral excited state (calculated at ca. 8.1 eV) can be transferred to molecular phonon modes \citep{marciniak21}. In this way, the system can storage enough vibrational energy to overcome the reaction barrier for fragmentation with selective C-C cleavage to multiple products \citep{los91,koster95}.

It is also to be noted that polymerization towards larger alkenes is not observed in our experiments (Sec.~\ref{sec:TPD}). This points out to a non-radical mediated photochemistry, opposite to what has been observed for shorter alkanes (e.g., CH$_4$, C$_2$H$_6$ and C$_3$H$_8$) \citep{carrascosa20}. In the latter case, it is well known that vacuum UV at low temperatures induces the formation of radicals by atomic hydrogen abstraction initiating a polymerization process. We did not observe this phenomenon and therefore the photochemistry of mid- to long alkanes seems different, supporting the photodissociation mechanism that we are proposing.

\section{Discussion}\label{sec:discussion}
As we have previously demonstrated that at the conditions of the circumstellar envelopes (CSEs) of C-rich Asymptotic Giant Branch (AGB) stars mainly aliphatics are produced \citep{martinez20}, plausible scenarios for the formation of aromatics needs to be explored due to their presence in interstellar environments. Indeed, benzene has not been yet identified in AGBs but it has been reported in protoplanetary nebulae (PPNe) \citep{cernicharo01}, which suggest a UV-driven transition from aliphatics to aromatics in an evolutionary context. 

On the other hand, the carbonaceous cosmic dust formed in AGBs consist in amorphous hydrocarbon particles comprised of sp$^2$ and sp$^3$ hybridizations \citep{Andersen2003}. The evolution from its formation in AGBs towards the PPNe and subsequent PNe phases increases the aromatic content to the detriment of the aliphatic portion \citep{Joblin1996, Goto2007}, which is consistent with the thermal annealing of the grains by the stochastic heating of energetic photons \citep{Goto2000}.

In the case of the diffuse ISM, dehydrogenation of the aliphatic portion of carbonaceous dust grains by the local interstellar UV radiation field is considered as the driving force towards aromatization \citep{jones12}, a process that will also occur in Photon-Dominated Regions (PDR) on estimated timescales of $\sim$ 10$^3$ yr or even lower \citep{jones12_corr}. UV-induced dehydrogenation in the ISM has been proposed to proceed through direct photodissociation of C-H bonds in aliphatic molecular species and in hydrogenated carbon grains \citep{caro01,mennella01,dartois05}. However, our results point towards a different scenario as we have shown that C-C photocleavage is favoured over C-H bond breaking. Therefore, hydrogen depletion in carbonaceous dust grains is primarily a result of the photolysis of aliphatic C-C bonds which produces small molecular fragments that might subsequently desorb through non-thermal processes \citep{Fredon2021,Dartois2022,DelFre2023} carrying hydrogen towards the gas phase and provoking an effective dehydrogenation of the carbon grains. 

It is also likely that the cleavage of the aliphatic C-C bonds of carbonaceous grains induces a structural rearrangement of the bond network towards an olefinic-rich material \citep{Smith1984,jones12}. According to our calculations the bond cleavage proceeds first by an elongation of the C-C bond and a subsequent H relocation. For aliphatic C-C bonds in a three dimensional carbon network such as HAC, the dangling C bond formed in the first step can interact with the network before H relocation occurs. This can liberate hydrogen from the carbonaceous material during the formation of the new bonding structure with the three dimensional network enabling the dissipation of any energy excess. This process can efficiently dehydrogenate the carbonaceous dust if, as suggested by our results, C-C cleavage is more prone than C-H bond breaking. Nevertheless, caution needs to be taken to directly extrapolate our results to complex carbon structures.  

The mechanism of C-C photocleavage that we describe here agrees with previous experimental observations on the UV processing of carbonaceous interstellar dust analogues in which in addition to H$_2$ production, the formation of alipahtic hydrocarbons (including olefins) up to four carbon atoms has been reported \citep{caro01, alata15}. The observation of C=C bonds of vinyl and \textit{trans}-vinylene nature has also been experimentally observed during the UV irradiation of several aliphatic hydrocarbons at conditions of the diffuse ISM \citep{dartois05}. Nevertheless, a thorough description of the UV-induced chemistry has not been provided despite its importance for modelling the chemical evolution of hydrocarbons and carbonaceous dust grains in the ISM. 

On the other hand, despite the fact that our results are general and not restricted to Dense Molecular Clouds (DMCs), they might contribute to explain the rich chemistry that has been recently identified in dark clouds. Small sized polycyclic aromatic hydrocarbons (PAHs) have been firmly identified in these environments \citep{mcguire21,cernicharo21,burkhardt21_b}, especifically in the Taurus Molecular Cloud (TMC-1), raising the question on how PAHs can be formed in these cold environments.

Chemical models have been able to satisfactorily reproduce the observed abundances of many of the more than 40 pure hydrocarbon species identified in TMC-1 but they have failed in explaining the abundances of cyclic molecules, systematically leading to lower values than those observed \citep{burkhardt21_b,mcguire21,mccarthy21_b}. At present it is still not clear if aromatics are formed through top-down or bottom-up processes. Top-down approaches have been suggested to be responsible for the large abundance of aromatics in TMC-1 \citep{burkhardt21_a},which might be inherited from a previous diffuse phase. On the other hand, the recent discovery of 1-cyano-1,3-butadiene (C$_4$H$_5$CN) \citep{cooke23} and the spatial distribution of C$_6$H$_5$CN \citep{cernicharo23} have been used to argue that the formation of aromatics proceeds through bottom-up chemical routes.

The most likely precursor for benzene formation in dark clouds from a bottom-up process is 1,3-butadiene (C$_4$H$_6$), which is known to lead to benzene (C$_6$H$_6$) through a barrierless and exoergic reaction with C$_2$H \citep{jones_kaiser11}. C$_4$H$_6$ has no permanent dipole moment and is therefore invisible at radiowavelengths, but the identification of C$_4$H$_5$CN  supports the presence of 1,3-butadiene in this environment \citep{morales11}. Thus, C$_4$H$_6$ might be a key species in the chemistry of TMC-1 and bottom-up chemical routes for the formation of aromatics in cold interstellar environments are dependent on efficient synthetic pathways for C$_4$H$_6$. 

A plausible formation of C$_4$H$_6$ involves the reaction of propene (C$_3$H$_6$) with the methylidyne (CH) radical \citep{daugey05, smith06, loison09}. C$_3$H$_6$ was detected more than 15 years ago in TMC-1 with fairly large abundance \citep{marcelino07}. However, to date no efficient reaction pathways towards C$_3$H$_6$ at low temperatures have been identified, suggesting that its formation is not driven by gas-phase chemistry \citep{lin13}.

In the evolution of cosmic dust grains in DMCs, a carbonaceous mantle is considered to be formed around dust grains by the accretion of gas-phase carbon atoms. Indeed, the C-C photocleavage process that we describe in detail here can operate in the photon dominated regions (PDRs) of molecular clouds \citep{pety05,alata2014,jones17} and our results show that UV-induced photocleavage of mid- to long alkanes or aliphatic moieties of carbonaceous dust grains lead to C$_3$H$_6$. Likewise, it provides a pathway for the formation of dienes in these environments by the UV photoprocessing of carbonaceous mantles that accrete on dust particles in the early stages of DMCs evolution \citep{jones13,murga23}.These newly formed mantles consist of aliphatic-rich material \citep{Ysard2015, Jones2016, murga23} and we speculate that if C$_3$H$_6$ and small dienes are photoformed on the surface of grains, they can desorb through non-thermal processes \citep{Fredon2021,Dartois2022,DelFre2023} to be incorporated into the gas-phase. The UV-induced formation of C$_3$H$_6$ and dienes might be therefore essential to understand the synthesis of small PAHs in dark clouds.

Finally, the mechanism that we have presented is also efficient in the formation of vinyl moieties, with important implications on the chemical formation routes of vinyl-bearing molecules in the interstellar medium. Vinyl containing Complex Organic Molecules (COMs) constitute an important, prevalent molecular class among the molecules detected in the ISM and DMCs, many of which have been detected in the last few years \citep{gardner75,hollis04,agundez21,cernicharo21_vynil,lee21,rivilla22,molpeceres22}.

\section{Conclusions}\label{sec:conclusions}
Our results explain in detail the photo-fragmentation mechanism of alkanes and alkane moieties in the ISM and show that C-C bond photocleavage is more likely than C-H bond breaking. The mechanism that we present here indicates that the dehydrogenation of dust towards an aromatic enrichment might proceed to a large extent as an effective process in which the fragments formed by the UV-photoprocessing of aliphatic-rich dust grains carry hydrogen towards the gas-phase after non-thermal desorption processes. This is particularly relevant to understand and more precisely modelling the evolution of cosmic dust among the different astrophysical environments.

Furthermore, we have observed an efficient formation of olefins, including propene and dienes. These have been suggested as plausible precursors for the gas-phase synthesis of aromatics in dark clouds, where gas-phase reactions are very much restricted. To explain the observed abundances of aromatics in these environments, propene and small dienes need to be incorporated in the very first steps of the chemical evolution models but, despite being detected, its formation mechanism is unknown which currently constitutes the main bottleneck for establishing consistent chemical routes towards the formation of aromatics in dark clouds. Although speculative, our findings might contribute to explain the high abundance of propene observed in TMC-1 and, by extension, of aromatics.

\begin{acknowledgments}
GT-C acknowledges funding from the Comunidad Aut\'{o}noma de Madrid (Grant No. PEJ-2021-AI/IND-21143). JIM-M acknowledges grant PID2021-125604NB-I00 MCIN/AEI/ 10.13039/501100011033 by the “European Union NextGenerationEU / PRTR”. GS acknowledges grant RYC2020-029810-I funded by MCIN/AEI/10.13039/501100011033 and by “ESF Investing in your future”. This work has been partially funded by grant PID2020-113142RB-C21 funded by MCIN/AEI/ 10.13039/501100011033 and grant PLEC2021-007906 funded by MCIN/AEI/ 10.13039/501100011033 by the “European Union NextGenerationEU/PRTR”. Partial funding by FotoArt-CM (P2018/NMT 4367) and Photosurf-CM (Y2020/NMT-6469) projects funded by Comunidad Aut\'{o}noma de Madrid and co-financed by European Structural Funds is also acknowledged.
\end{acknowledgments}

\newpage

\appendix
\section{IR band assignment}\label{app_assign}

Table~\ref{table:IR_assignment} lists the most prominent absorption features of amorphous C$_6$H$_{14}$, crystalline C$_6$H$_{14}$ and C$_{11}$H$_{24}$ at low temperature along with its vibrational assignment.
\begin{table}[hbt!]
    \caption{Infrared band assignment}          
    \label{table:IR_assignment}      
    \centering
    \begin{tabular}{c c c l}
        \hline\hline
        \multicolumn{3}{c}{Wavenumber   (cm$^{-1}$)} & Assignment$^{(a),(b)}$ \\
        \multicolumn{2}{c}{C$_6$H$_{14}$}  & C$_{11}$H$_{24}$     &                      \\
        \hline
        Amorphous   & Crystalline  & Amorphous  &                      \\
        2961        & 2962         & 2963       & $\nu_{as,ip}$ CH$_3$           \\
        2956        & 2952         & 2957       & $\nu_{as,oop}$ CH$_3$          \\
        2930        & 2929         & 2932       & $\nu_{s,F}$ CH$_2$             \\
        2923        & 2916         & 2924       & $\nu_{as}$ CH$_2$              \\
        2898        & 2896         & 2901       & $\nu_{s,F}$ CH$_2$             \\
        2870        & 2869         & 2870       & $\nu_s$ CH$_3$               \\
        2857        & 2853         & 2855       & $\nu_{s,F}$ CH$_2$             \\
        2850        & 2847         & 2849       & $\nu_{s,F}$ CH$_2$             \\
        1470        & 1473         & 1468       & $\delta_b$ CH$_2$               \\
        1458        & 1461         & 1459       & $\delta_{sc}$ CH$_2$, $\delta_{as}$ CH$_3$   \\
        1440        & 1450         & 1442       & $\delta_{sc}$ CH$_2$, $\delta_{as}$ CH$_3$ \\
        1377        & 1367         & 1378       & $\delta_{s}$ CH$_3$               \\
        1341        & –            & –          & $\gamma$ CH$_2$              \\
        –           & 1065         & –          & $\nu$ C-C                \\
        –           & 885          & –          & $\rho$ CH$_3$               
    \end{tabular}
    \tablecomments{$^{(a)}$The vibrational modes are abbreviated as follows: $\nu$: stretching; $\delta$: deformation (b: bend; sc: scissor); $\gamma$: wagging; $\rho$: rocking; \textit{s}: symmetric; \textit{as}: asymmetric; \textit{ip}: in-plane; \textit{oop}: out-of-plane. $^{(b)}$Assignments from \cite{snyder63,snyder78,jordanov03}.
    }
\end{table}
\newpage

\section{Quantitative analysis of the IR spectra of C$_6$H$_{14}$}\label{sec:app_fit}

\begin{figure}[hbt!]
    \centering
    \includegraphics[width=0.75\textwidth]{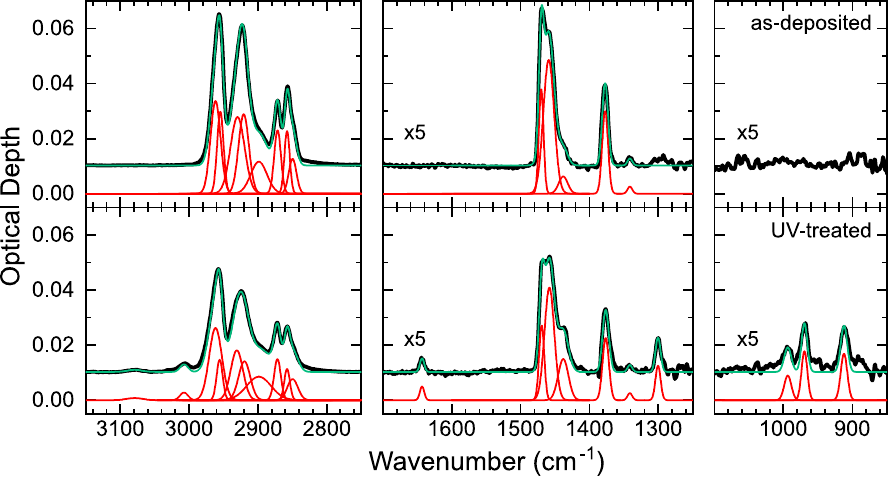}
    \caption{Spectral fitting of as-deposited C$_6$H$_{14}$ (top) and after a UV fluence of 9.8 $\times$ 10$^{18}$ ph cm$^{-2}$ (bottom). Experimental spectra: black; Gaussian profiles: red; Fitting result: green. The experimental spectra and the fitting results are shifted vertically for clarity.}
    \label{fig:app_fit}
\end{figure}

To derive the effective cross-sections of photodestruction, $\sigma_{des}$, and photoformation, $\sigma_{form}$, for several molecular moieties, we performed a quantitative analysis of the IR spectra acquired during UV irradiation by fitting selected IR absorption features assuming Gaussian profiles. For the fitting of the CH$_2$/CH$_3$ stretching mode region (2800-300 cm$^{-1}$) we used eight Gaussian curves according to \cite{snyder78} and \cite{jordanov03} to account for the complex structure due to Fermi resonances. During the fitting process, peak positions were allowed to vary by $\pm$1 cm$^{-1}$, whereas peak full widths at half maximum (FWHM) were restricted to values lower than 30 cm$^{-1}$, except for the case of the $\nu_{s,F}$ CH$_2$ mode at 2898 cm$^{-1}$ that was restricted to values lower than 50 cm$^{-1}$. The increase in FWHM of this last peak upon UV irradiation is the reason for relaxing the fitting constrain. This increase is related to the new molecules that are formed, whose spectral response changes the Fermi resonances of the CH$_2$ bending overtones with the fundamental CH$_2$ stretching modes. Figure~\ref{fig:app_fit} shows the fitting results of the IR spectra for as-deposited C$_6$H$_{14}$ and after the complete UV treatment.
\newpage

\section{IR spectra of amorphous C$_6$H$_{14}$ during UV irradiation}\label{app:IR_spectra}

\begin{figure}[hbt!]
    \centering
    \includegraphics[width=0.75\textwidth]{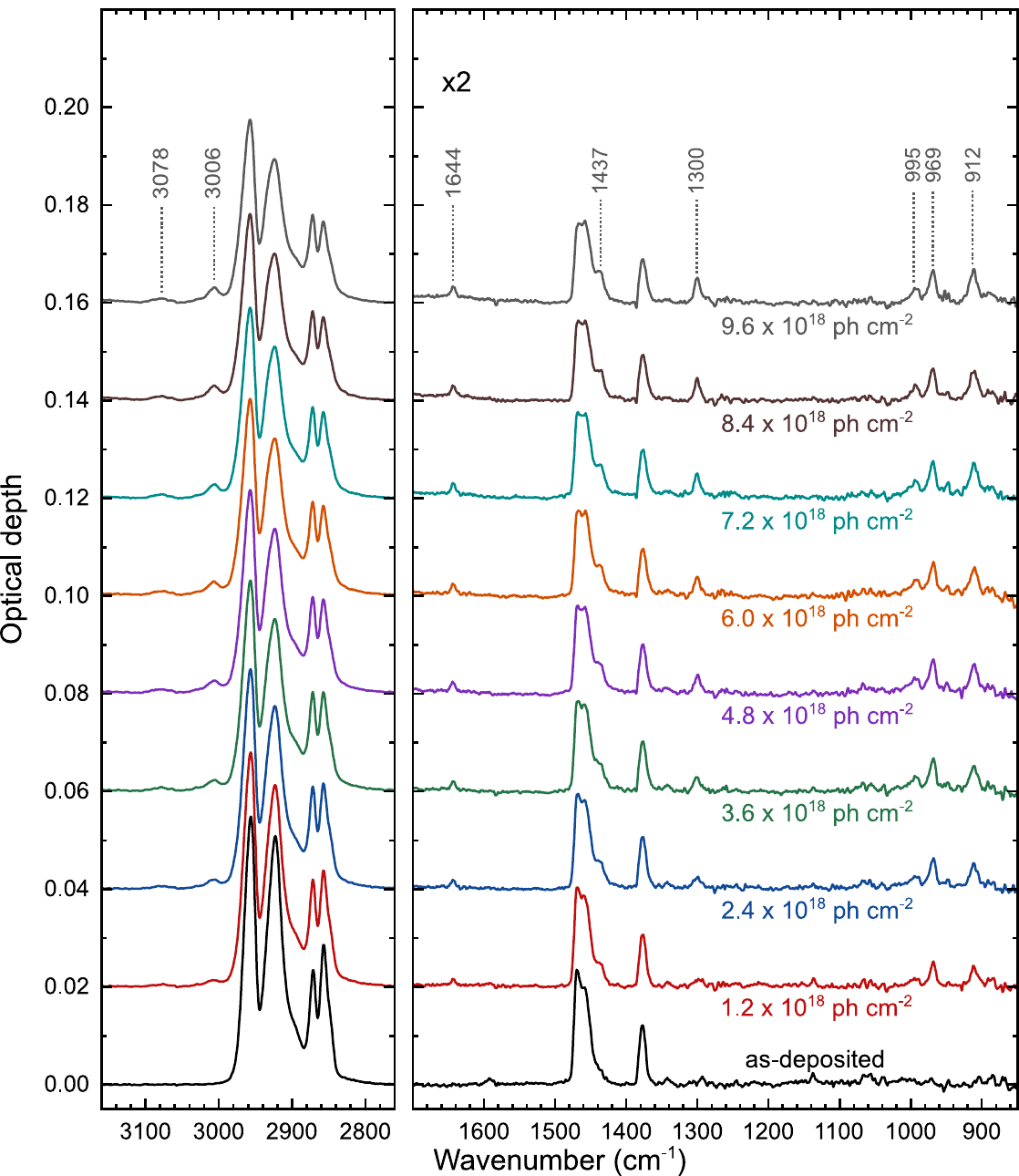}
    \caption{IR spectra of amorphous C$_6$H$_{14}$ at different UV fluences. The new IR bands upon UV exposure are indicated as well as the corresponding UV fluence of each spectrum. The spectra are vertically shifted for clarity.}
    \label{fig:appam_C6H14}
\end{figure}

Figure~\ref{fig:appam_C6H14} shows selected IR spectra obtained during UV irradiation of C$_6$H$_{14}$ at 14 K. The emergence of new IR features can be observed along with the decrease of the bands associated to C$_6$H$_{14}$.
\newpage

\section{UV irradiation of crystalline C$_6$H$_{14}$}\label{app:cristC6H14}

\begin{figure}[hbt!]
    \centering
    \includegraphics[width=0.75\textwidth]{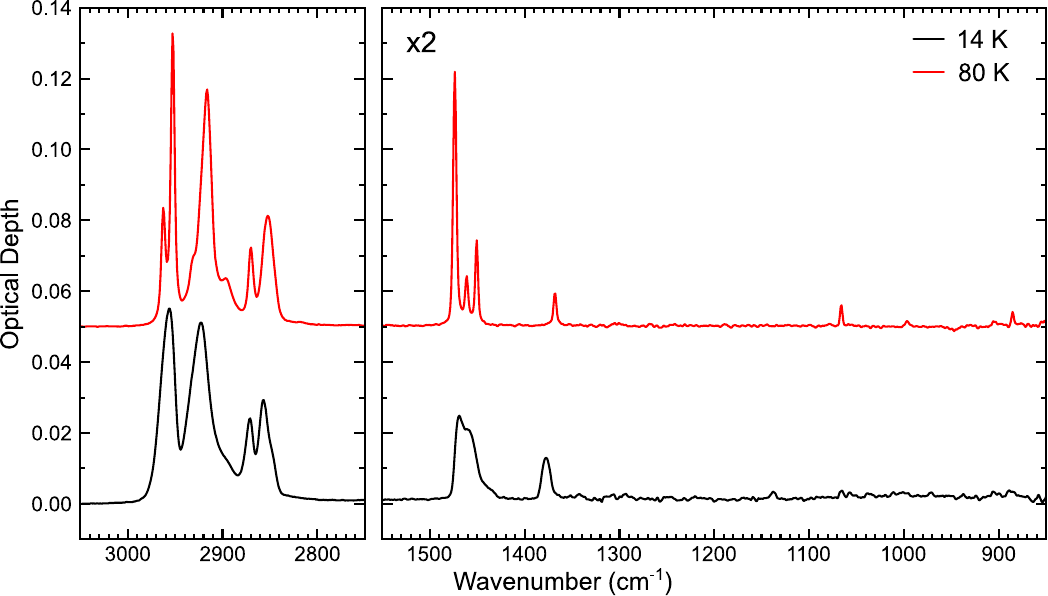}
    \caption{IR spectra of crystalline (red; deposition temperature: 80 K) and amorphous (black; deposition temperature: 14 K) solid C$_6$H$_{14}$. The spectra are shifted vertically for clarity.}
    \label{fig:app_crist_am_C6H14}
\end{figure}

The morphology of solid C$_6$H$_{14}$ films depends on the substrate temperature during deposition. Whilst at 14 K an amorphous solid is obtained, a substrate temperature of 80 K produces a crystalline one, as revealed by the IR spectra (Fig.~\ref{fig:app_crist_am_C6H14})
\newpage

\begin{figure}[hbt!]
    \centering
    \includegraphics[width=0.75\textwidth]{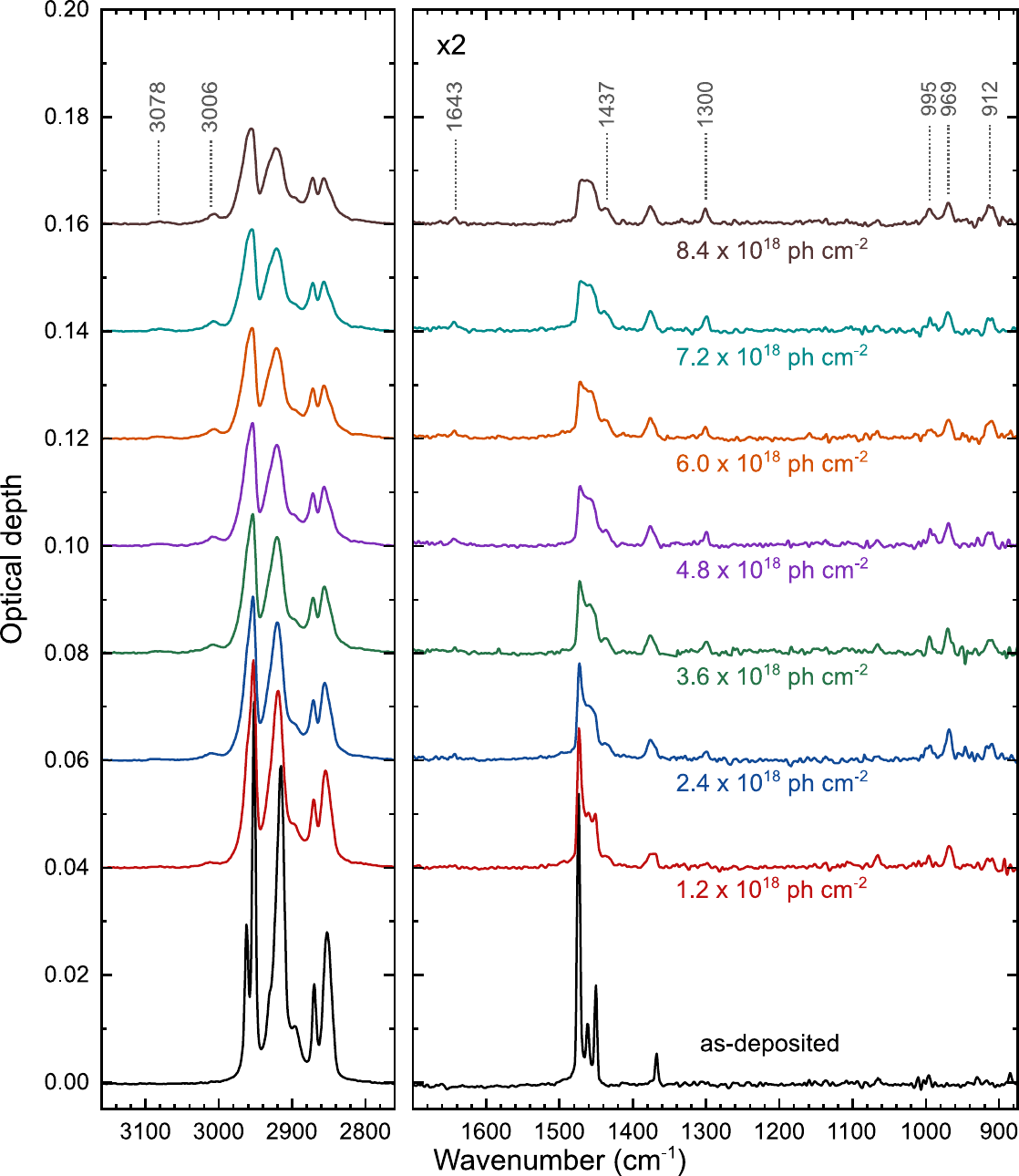}
    \caption{IR spectra of crystalline C$_6$H$_{14}$  at different UV fluences. The new IR bands upon UV exposure are indicated as well as the corresponding UV fluence of each spectrum. The spectra are shifted vertically for clarity.}
    \label{fig:app_crist_C6H14}
\end{figure}

The UV irradiation of crystalline C$_6$H$_{14}$ induces the same photochemistry as in the case of amorphous C$_6$H$_{14}$, i.e., the formation of olefinic moieties both of vinyl (-CH=CH$_2$) and \textit{trans}-vinylene (-CH=CH-) nature along with the formation of CH$_4$ due to the photocleavage of the alkane C-C bonds, which, during the first stages of the UV irradiation, disrupts the crystalline morphology (Fig.~\ref{fig:app_crist_C6H14}).

On the other hand, as stated in Section~\ref{sec:IR}, the relationship of the band at 1437 cm$^{-1}$ with the formation of C=C bonds becomes more evident in the case of crystalline C$_6$H$_{14}$, since no overlapping of the CH$_2$/CH$_3$ deformation modes at about 1480-1440 cm$^{-1}$ occurs in the first IR spectra during UV exposure. This supports our assignment of this band to the deformation of methylene groups in the presence of adjacent unsaturated groups (Table~\ref{table:ir}).
\newpage

\section{UV irradiation of C$_{11}$H$_{24}$}\label{app:C11H24}

As mentioned in Section~\ref{sec:results}, the mechanism of formation of olefins through UV irradiation is not restricted to C$_6$H$_{14}$ but general to mid- and long-chain linear alkanes. 

\begin{figure}[hbt!]
    \centering
    \includegraphics[width=0.75\textwidth]{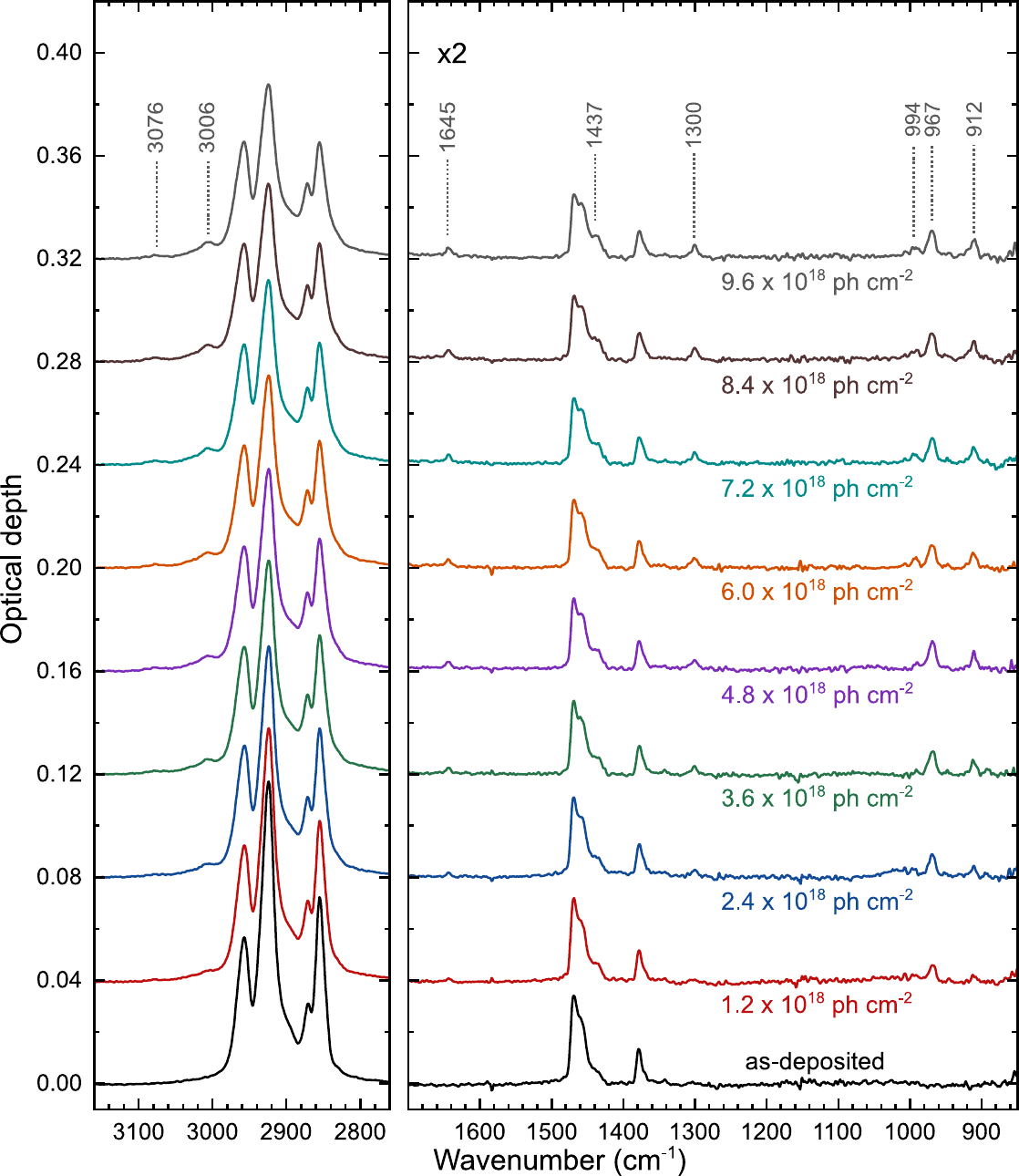}
    \caption{IR spectra of amorphous C$_{11}$H$_{24}$ at different UV fluences. The new IR bands upon UV exposure are indicated as well as the corresponding UV fluence of each spectrum. The spectra are shifted vertically for clarity.}
    \label{fig:app_c11h24}
\end{figure}

To illustrate this, Figure ~\ref{fig:app_c11h24} shows the IR spectra of amorphous undecane (C$_{11}$H$_{24}$) at different irradiation fluences. The formation of –CH=CH– and –CH=CH$_2$ moieties and CH$_4$ is evident from the spectra. The band assignment of the new IR bands corresponds to those listed in Section~\ref{sec:IR}  for C$_6$H$_{14}$ (Table~\ref{table:ir}).
\newpage

\section{Excited electronic states of C$_6$H$_{14}$}\label{app:excited_states}

\begin{figure}[hbt!]
    \centering
    \includegraphics[width=0.5\textwidth]{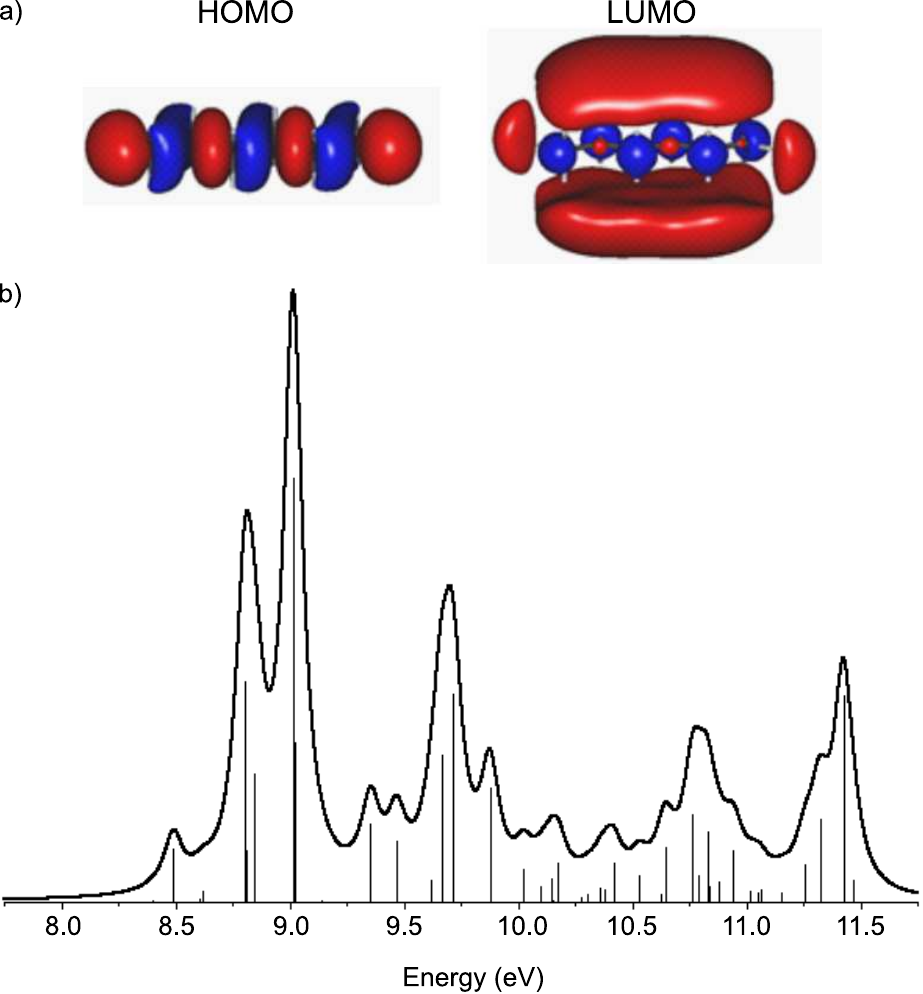}
    \caption{a) Isosurface of the molecular orbitals involved in the transition of the first excited state S$_1$. b) Simulated absorption spectra of neutral C$_6$H$_{14}$ vertical excitation at the relaxed ground state structure (CAM-B3LYP/6-31++G*): transitions (bars) fitted with a Lorentzian lineshape (FWHM = 0.05eV; solid line).}
    \label{fig:HOMO}
\end{figure}

Figure~\ref{fig:HOMO}a shows the isosurface of the HOMO and LUMO molecular orbitals of C$_6$H$_{14}$. It can be shown that the LUMO is highly delocalized which is consistent with the Rydberg character of the orbital. In addition, in Table~\ref{table:HOMO} we have listed the vertical excitation energy and weight of the HOMO $\rightarrow$ LUMO transition which is the  dominant contribution to the S$_1$ excited state. The same excitation with Rydberg character also dominates the first triplet excited state (T1) at 8.01 eV.

We have also simulated the absorption spectrum of neutral C$_6$H$_{14}$ from the calculations of the transitions from 8 eV to 11.5 eV. The results are shown in Figure~\ref{fig:HOMO}b.

\begin{deluxetable}{r c c c}[hbt!]
    \label{table:HOMO}
    \tablecaption{Vertical excitation energy, weight of the dominant transition and character of the transition for the  first excited state S$_1$ (CAM-B3LYP/6-31++G* calculations).}
    \tablewidth{\columnwidth}
    \tablehead{
    \colhead{} & \colhead{Energy} & \colhead{Weight} & \colhead{Transition}\\
    \colhead{} & \colhead{eV (nm)}}

        \startdata
        S$_1$    & 8.09 (153.3) & 0.87 & HOMO $\rightarrow$ LUMO\\
        \enddata

\end{deluxetable}
\newpage

\section{Energy barriers for C-C photocleavage}\label{app:barriers}

\begin{figure}[hbt!]
    \centering
    \includegraphics[width=0.6 \textwidth]{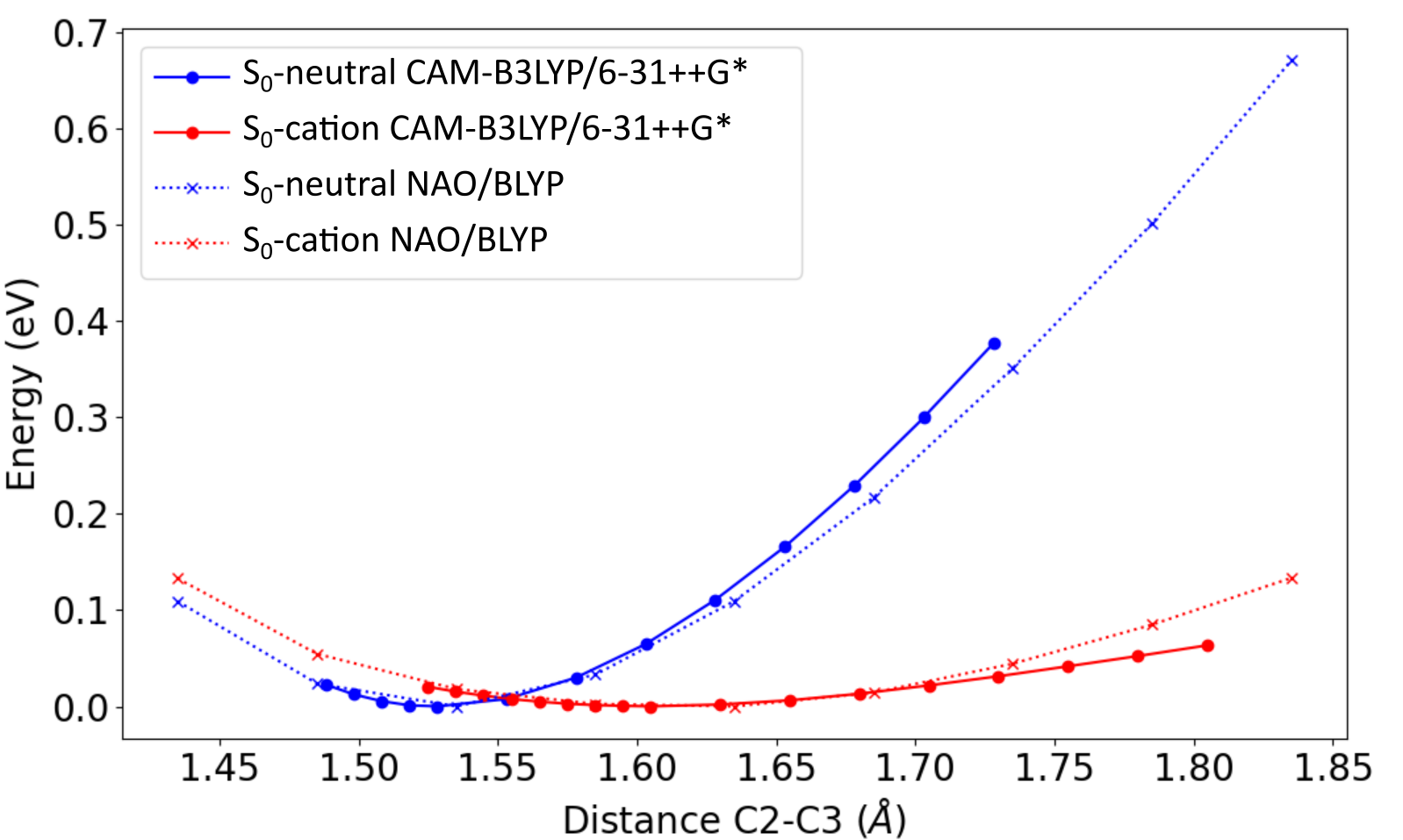}
    \caption{Energy profiles for C2-C3 distance around the minimum energy structure for S0 and cation states with different DFT approximations}
    \label{fig:DFT_comparison}
\end{figure}

For scanning the conformational space of the reactions we have chosen the BLYP functional given its capability for studying a larger conformational space. To validate its use in Figure~\ref{fig:DFT_comparison} we show a comparison of the energy profiles around the minimum energy structure for the S$_0$-neutral and S$_0$-cation states using different DFT approximations, namely CAM-B3LYP/6-31++G* using Gaussian 16 and BLYP-D3/NAO using Fireball. The agreement is very good validating our selection of the BLYP functiontal.

\begin{figure}[hbt!]
    \centering
    \includegraphics[width=0.69 \textwidth]{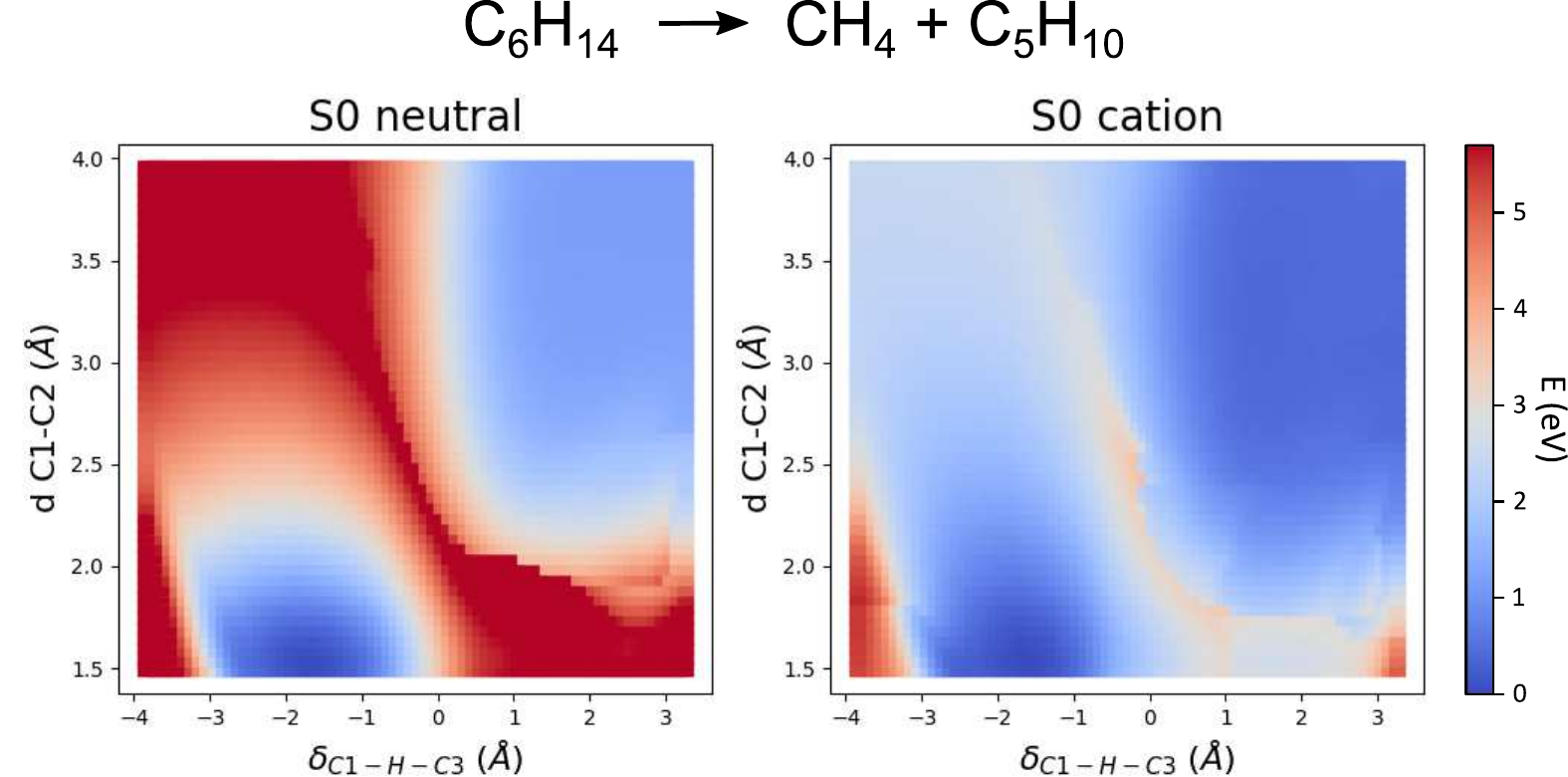}
    \caption{Energy landscapes for the photocleavage reaction C$_6$H$_{14}$ $\rightarrow$ CH$_4$ + C$_5$H$_{10}$. The reactions coordinates are $\delta$=d(C1-H)-d(C3-H) vs. the distance between C1 and C2.}
    \label{fig:app_barriers_CH4}
\end{figure}

\begin{figure}[hbt!]
    \centering
    \includegraphics[width=0.69 \textwidth]{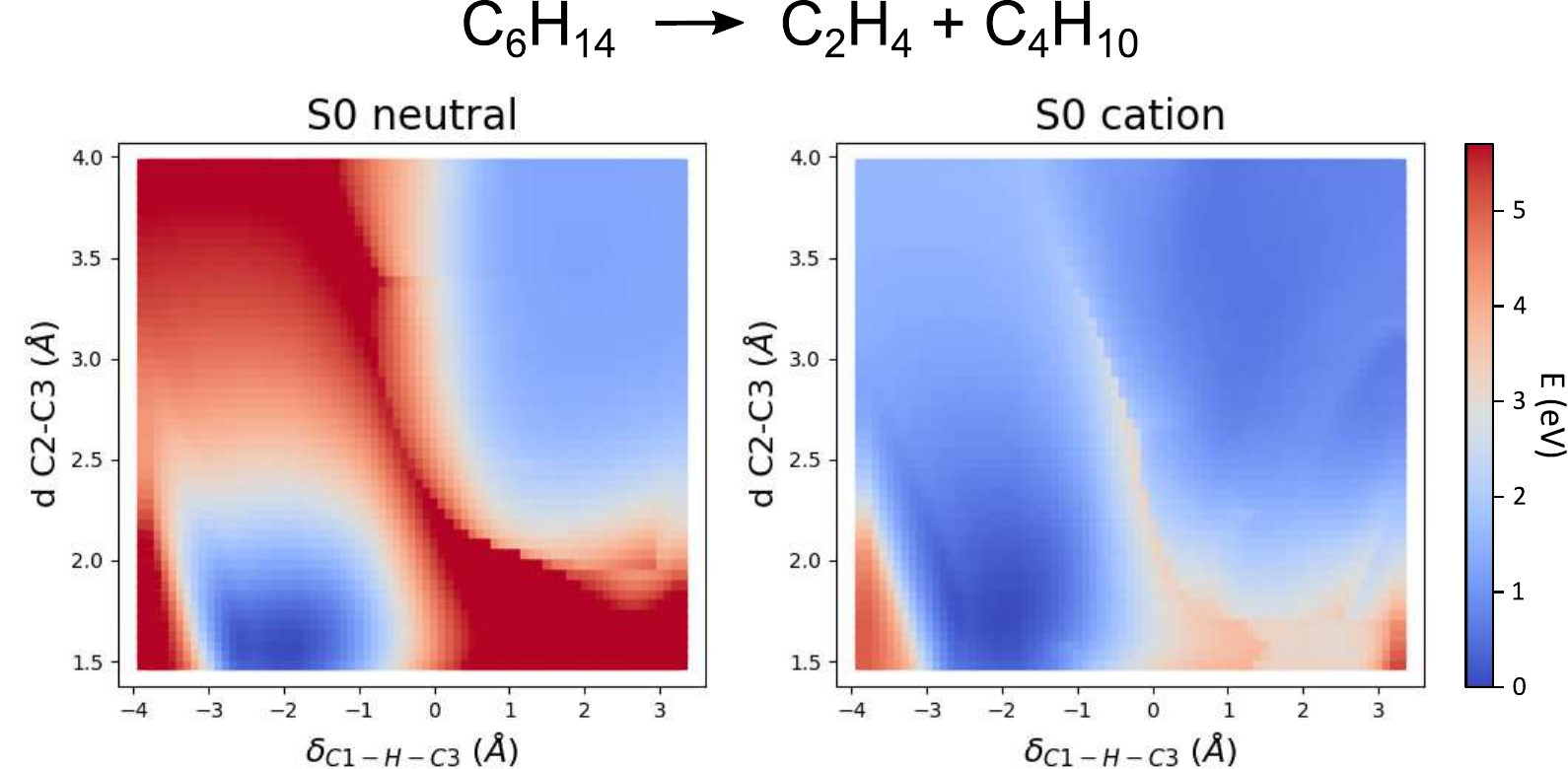}
    \caption{Energy landscapes for the photocleavage reaction C$_6$H$_{14}$ $\rightarrow$ C$_2$H$_4$ + C$_4$H$_{10}$. The reactions coordinates are $\delta$=d(C3-H)-d(C1-H) vs. the distance between C2 and C3.}
    \label{fig:app_barriers_C2H4}
\end{figure}

\begin{figure}[hbt!]
    \centering
    \includegraphics[width=0.69 \textwidth]{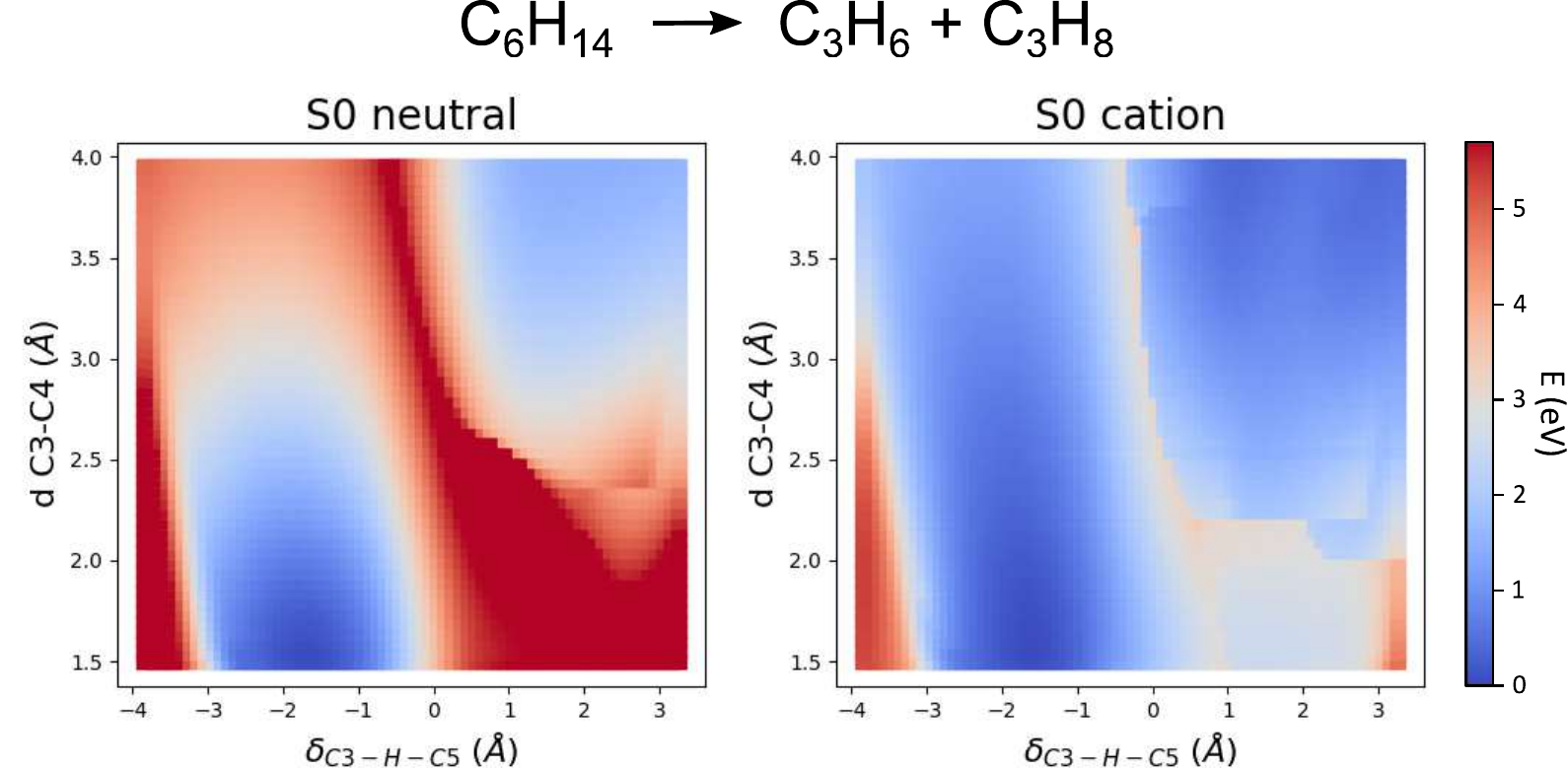}
    \caption{Energy landscapes for the photocleavage reaction C$_6$H$_{14}$ $\rightarrow$ C$_3$H$_6$ + C$_3$H$_8$. The reactions coordinates are ($\delta$=d(C3-H)-d(C5-H) vs. the distance between C3 and C4.}
    \label{fig:app_barriers_C3H6}
\end{figure}

\begin{figure}[hbt!]
    \centering
    \includegraphics[width=0.69 \textwidth]{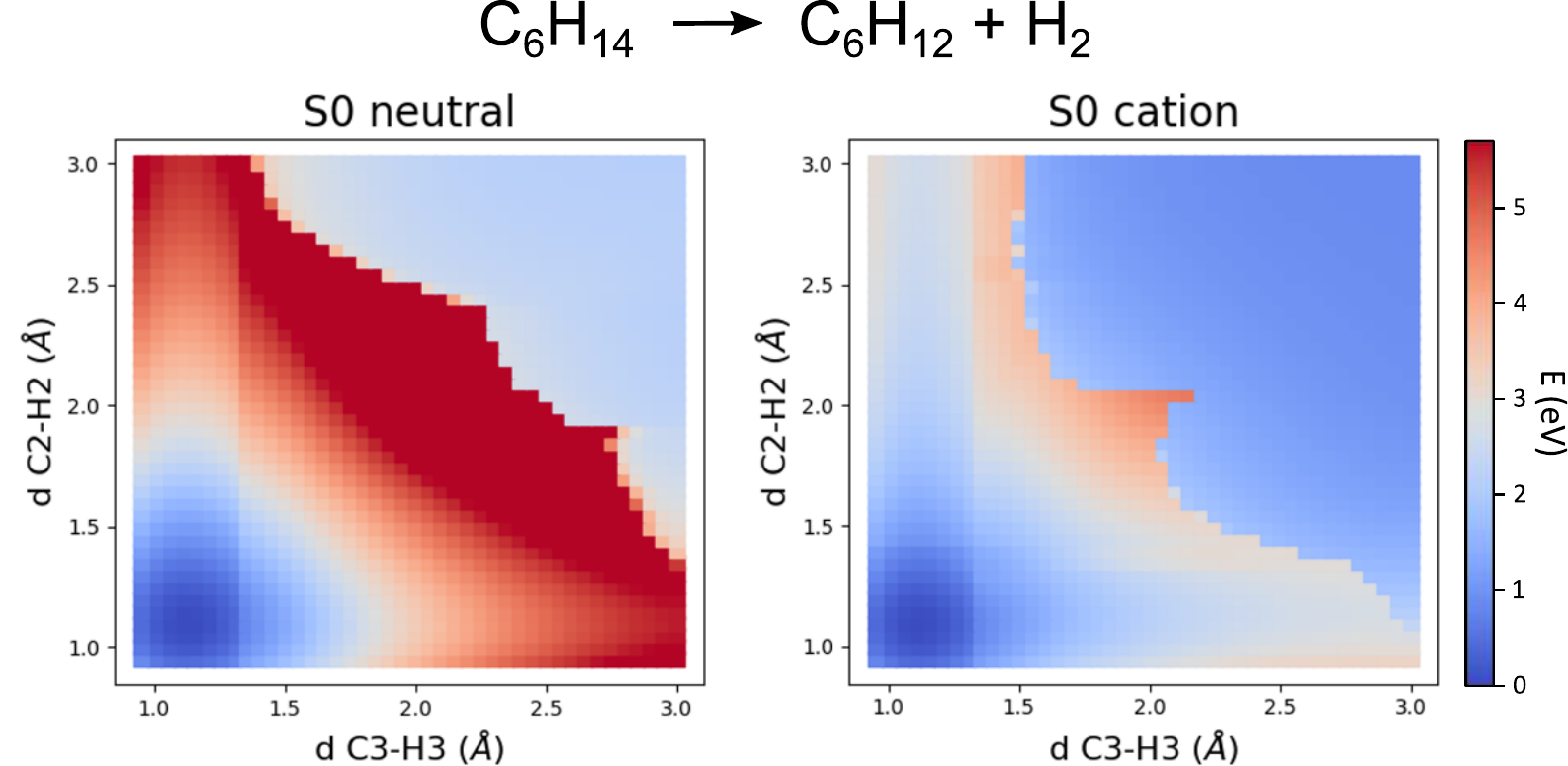}
    \caption{Energy landscapes for the photocleavage reaction C$_6$H$_{14}$ $\rightarrow$ C$_6$H$_{12}$ + H$_2$. The reactions coordinates are the distance between the H and the C in C2 and the distance and between the H and the C in C3.}
    \label{fig:app_barriers_H2}
\end{figure}

In addition to the results shown in Section~\ref{sec:calc} for the energy barriers for the reaction C$_6$H$_{14}$ $\rightarrow$ C$_2$H$_6$ + C$_4$H$_8$ (C2-C3 photocleavage), we have explored several other reactions in which cleavage of other C-C bonds is involved. We have also investigated the reaction leading to H$_2$ elimination through C-H cleavage. We note that all the barriers for C-C cleavage are lower than that for C-H bond breaking (both in the neutral and cationic ground states) which corroborates that C-C cleavage is a more probable process upon UV excitation. The energy landscapes for the explored reactions are presented in Figs.~\ref{fig:app_barriers_CH4},~\ref{fig:app_barriers_C2H4},~\ref{fig:app_barriers_C3H6} and~\ref{fig:app_barriers_H2}.

\newpage

\bibliography{Tajuelo-Castilla_C6H14.bib}{}

\begin{thebibliography}{}
\expandafter\ifx\csname natexlab\endcsname\relax\def\natexlab#1{#1}\fi
\providecommand{\url}[1]{\href{#1}{#1}}
\providecommand{\dodoi}[1]{doi:~\href{http://doi.org/#1}{\nolinkurl{#1}}}
\providecommand{\doeprint}[1]{\href{http://ascl.net/#1}{\nolinkurl{http://ascl.net/#1}}}
\providecommand{\doarXiv}[1]{\href{https://arxiv.org/abs/#1}{\nolinkurl{https://arxiv.org/abs/#1}}}

\bibitem[{Accolla {et~al.}(2021)Accolla, Santoro, Merino, Mart{\'{\i}}nez, Tajuelo-Castilla, V{\'{a}}zquez, Sobrado, Ag{\'{u}}ndez, Jim{\'{e}}nez-Redondo, Herrero, Tanarro, Cernicharo, \& Mart{\'{\i}}n-Gago}]{accolla21}
Accolla, M., Santoro, G., Merino, P., {et~al.} 2021, The Astrophysical Journal, 906, 44, \dodoi{10.3847/1538-4357/abc703}

\bibitem[{Ag{\'{u}}ndez {et~al.}(2021)Ag{\'{u}}ndez, Marcelino, Tercero, Cabezas, de~Vicente, \& Cernicharo}]{agundez21}
Ag{\'{u}}ndez, M., Marcelino, N., Tercero, B., {et~al.} 2021, Astronomy {\&} Astrophysics, 649, L4, \dodoi{10.1051/0004-6361/202140978}

\bibitem[{{Alata} {et~al.}(2014){Alata}, {Cruz-Diaz}, {Mu{\~n}oz Caro}, \& {Dartois}}]{alata2014}
{Alata}, I., {Cruz-Diaz}, G.~A., {Mu{\~n}oz Caro}, G.~M., \& {Dartois}, E. 2014, A\&A, 569, A119, \dodoi{10.1051/0004-6361/201323118}

\bibitem[{Alata {et~al.}(2015)Alata, Jallat, Gavilan, Chabot, Cruz-Diaz, Caro, B{\'{e}}roff, \& Dartois}]{alata15}
Alata, I., Jallat, A., Gavilan, L., {et~al.} 2015, Astronomy {\&} Astrophysics, 584, A123, \dodoi{10.1051/0004-6361/201526368}

\bibitem[{{Allamandola} {et~al.}({1985}){Allamandola}, {Tielens}, \& {Barker}}]{allamandola85}
{Allamandola}, L., {Tielens}, A., \& {Barker}, J. {1985}, The Astrophysical Journal, 290, L25, \dodoi{{10.1086/184435}}

\bibitem[{{Allamandola} {et~al.}(1989){Allamandola}, {Tielens}, \& {Barker}}]{allamandola89}
---. 1989, The Astrophysical Journal Supplement series, 71, 733, \dodoi{{10.1086/191396}}

\bibitem[{{Andersen} {et~al.}(2003){Andersen}, {H{\"o}fner}, \& {Gautschy-Loidl}}]{Andersen2003}
{Andersen}, A.~C., {H{\"o}fner}, S., \& {Gautschy-Loidl}, R. 2003, \aap, 400, 981, \dodoi{10.1051/0004-6361:20030036}

\bibitem[{Basanta {et~al.}(2007)Basanta, Dappe, Jel{\'{\i}}nek, \& Ortega}]{basanta07}
Basanta, M., Dappe, Y., Jel{\'{\i}}nek, P., \& Ortega, J. 2007, Computational Materials Science, 39, 759, \dodoi{10.1016/j.commatsci.2006.09.003}

\bibitem[{Bernstein {et~al.}(2002)Bernstein, Dworkin, Sandford, Cooper, \& Allamandola}]{bernstein02}
Bernstein, M.~P., Dworkin, J.~P., Sandford, S.~A., Cooper, G.~W., \& Allamandola, L.~J. 2002, Nature, 416, 401, \dodoi{10.1038/416401a}

\bibitem[{Burkhardt {et~al.}(2021{\natexlab{a}})Burkhardt, Loomis, Shingledecker, Lee, Remijan, McCarthy, \& McGuire}]{burkhardt21_a}
Burkhardt, A.~M., Loomis, R.~A., Shingledecker, C.~N., {et~al.} 2021{\natexlab{a}}, Nature Astronomy, 5, 181, \dodoi{10.1038/s41550-020-01253-4}

\bibitem[{Burkhardt {et~al.}(2021{\natexlab{b}})Burkhardt, Lee, Changala, Shingledecker, Cooke, Loomis, Wei, Charnley, Herbst, McCarthy, \& McGuire}]{burkhardt21_b}
Burkhardt, A.~M., Lee, K. L.~K., Changala, P.~B., {et~al.} 2021{\natexlab{b}}, The Astrophysical Journal Letters, 913, L18, \dodoi{10.3847/2041-8213/abfd3a}

\bibitem[{Caro {et~al.}(2002)Caro, Meierhenrich, Schutte, Barbier, Segovia, Rosenbauer, Thiemann, Brack, \& Greenberg}]{munozcaro02}
Caro, G., Meierhenrich, U., Schutte, W., {et~al.} 2002, Nature, 416, 403, \dodoi{10.1038/416403a}

\bibitem[{Carrascosa {et~al.}(2020)Carrascosa, Cruz-D{\'{\i}}az, Caro, Dartois, \& Chen}]{carrascosa20}
Carrascosa, H., Cruz-D{\'{\i}}az, G.~A., Caro, G. M.~M., Dartois, E., \& Chen, Y.-J. 2020, Monthly Notices of the Royal Astronomical Society, 493, 821, \dodoi{10.1093/mnras/staa334}

\bibitem[{Cecchi-Pestellini \& Aiello(1992)}]{CecchiPestellini92}
Cecchi-Pestellini, C., \& Aiello, S. 1992, Monthly Notices of the Royal Astronomical Society, 258, 125, \dodoi{10.1093/mnras/258.1.125}

\bibitem[{{Cernicharo} {et~al.}(2021){Cernicharo}, {Ag{\'u}ndez}, {Cabezas}, {Tercero}, {Marcelino}, {Pardo}, \& {de Vicente}}]{cernicharo21}
{Cernicharo}, J., {Ag{\'u}ndez}, M., {Cabezas}, C., {et~al.} 2021, \aap, 649, L15, \dodoi{10.1051/0004-6361/202141156}

\bibitem[{Cernicharo {et~al.}(2001)Cernicharo, Heras, Tielens, Pardo, Herpin, Gu{\'{e}}lin, \& Waters}]{cernicharo01}
Cernicharo, J., Heras, A.~M., Tielens, A. G. G.~M., {et~al.} 2001, The Astrophysical Journal, 546, L123, \dodoi{10.1086/318871}

\bibitem[{Cernicharo {et~al.}(2023)Cernicharo, Tercero, Marcelino, Ag{\'{u}}ndez, \& de~Vicente}]{cernicharo23}
Cernicharo, J., Tercero, B., Marcelino, N., Ag{\'{u}}ndez, M., \& de~Vicente, P. 2023, Astronomy {\&} Astrophysics, 674, L4, \dodoi{10.1051/0004-6361/202346722}

\bibitem[{Cernicharo {et~al.}(2021)Cernicharo, Ag{\'{u}}ndez, Cabezas, Marcelino, Tercero, Pardo, Gallego, Tercero, L{\'{o}}pez-P{\'{e}}rez, \& de~Vicente}]{cernicharo21_vynil}
Cernicharo, J., Ag{\'{u}}ndez, M., Cabezas, C., {et~al.} 2021, Astronomy {\&} Astrophysics, 647, L2, \dodoi{10.1051/0004-6361/202140434}

\bibitem[{Chen {et~al.}(2013)Chen, Chuang, Caro, Nuevo, Chu, Yih, Ip, \& Wu}]{chen13}
Chen, Y.-J., Chuang, K.-J., Caro, G. M.~M., {et~al.} 2013, The Astrophysical Journal, 781, 15, \dodoi{10.1088/0004-637x/781/1/15}

\bibitem[{Chevance {et~al.}(2019)Chevance, Kruijssen, Hygate, Schruba, Longmore, Groves, Henshaw, Herrera, Hughes, Jeffreson, Lang, Leroy, Meidt, Pety, Razza, Rosolowsky, Schinnerer, Bigiel, Blanc, Emsellem, Faesi, Glover, Haydon, Ho, Kreckel, Lee, Liu, Querejeta, Saito, Sun, Usero, \& Utomo}]{chevance19}
Chevance, M., Kruijssen, J. M.~D., Hygate, A. P.~S., {et~al.} 2019, Monthly Notices of the Royal Astronomical Society, 493, 2872, \dodoi{10.1093/mnras/stz3525}

\bibitem[{Chiar {et~al.}(2013)Chiar, Tielens, Adamson, \& Ricca}]{chiar13}
Chiar, J.~E., Tielens, A. G. G.~M., Adamson, A.~J., \& Ricca, A. 2013, ASTROPHYSICAL JOURNAL, 770, \dodoi{10.1088/0004-637X/770/1/78}

\bibitem[{Ciesla \& Sandford(2012)}]{ciesla12}
Ciesla, F.~J., \& Sandford, S.~A. 2012, Science, 336, 452, \dodoi{10.1126/science.1217291}

\bibitem[{Cooke {et~al.}(2023)Cooke, Xue, Changala, Shay, Byrne, Tang, Fried, Lee, Loomis, Lamberts, Remijan, Burkhardt, Herbst, McCarthy, \& McGuire}]{cooke23}
Cooke, I.~R., Xue, C., Changala, P.~B., {et~al.} 2023, The Astrophysical Journal, 948, 133, \dodoi{10.3847/1538-4357/acc584}

\bibitem[{{Cottin} {et~al.}(2003){Cottin}, {Moore}, \& {B{\'e}nilan}}]{cottin03}
{Cottin}, H., {Moore}, M.~H., \& {B{\'e}nilan}, Y. 2003, \apj, 590, 874, \dodoi{10.1086/375149}

\bibitem[{{Dartois} {et~al.}(2004){Dartois}, {Marco}, {Mu{\~n}oz-Caro}, {Brooks}, {Deboffle}, \& {d'Hendecourt}}]{dartois04}
{Dartois}, E., {Marco}, O., {Mu{\~n}oz-Caro}, G.~M., {et~al.} 2004, \aap, 423, 549, \dodoi{10.1051/0004-6361:20047067}

\bibitem[{{Dartois} {et~al.}(2005){Dartois}, {Mu{\~n}oz Caro}, {Deboffle}, {Montagnac}, \& {D'Hendecourt}}]{dartois05}
{Dartois}, E., {Mu{\~n}oz Caro}, G.~M., {Deboffle}, D., {Montagnac}, G., \& {D'Hendecourt}, L. 2005, \aap, 432, 895, \dodoi{10.1051/0004-6361:20042094}

\bibitem[{{Dartois} {et~al.}(2022){Dartois}, {Chabot}, {Koch}, {Bachelet}, {Bender}, {Bour{\c{c}}ois}, {Duprat}, {Frereux}, {Godard}, {Herv{\'e}}, {Merk}, {Pino}, {Rojas}, {Schubert}, \& {Trautmann}}]{Dartois2022}
{Dartois}, E., {Chabot}, M., {Koch}, F., {et~al.} 2022, \aap, 663, A25, \dodoi{10.1051/0004-6361/202243274}

\bibitem[{Daugey {et~al.}(2005)Daugey, Caubet, Retail, Costes, Bergeat, \& Dorthe}]{daugey05}
Daugey, N., Caubet, P., Retail, B., {et~al.} 2005, Physical Chemistry Chemical Physics, 7, 2921, \dodoi{10.1039/b506096f}

\bibitem[{de~Koster \& Beijersbergen(1995)}]{koster95}
de~Koster, C.~G., \& Beijersbergen, J. H.~M. 1995, Rapid Communications in Mass Spectrometry, 9, 1115, \dodoi{10.1002/rcm.1290091207}

\bibitem[{Del~Fr\'e {et~al.}(2023)Del~Fr\'e, Santamar\'{\i}a, Duflot, Basalg\`ete, F\'eraud, Bertin, Fillion, \& Monnerville}]{DelFre2023}
Del~Fr\'e, S., Santamar\'{\i}a, A.~R., Duflot, D., {et~al.} 2023, Phys. Rev. Lett., 131, 238001, \dodoi{10.1103/PhysRevLett.131.238001}

\bibitem[{Duley(2000)}]{Duley2000}
Duley, W.~W. 2000, The Astrophysical Journal, 528, 841, \dodoi{10.1086/308204}

\bibitem[{{Duley} \& {Williams}(1988)}]{Duley1988}
{Duley}, W.~W., \& {Williams}, D.~A. 1988, \mnras, 231, 969, \dodoi{10.1093/mnras/231.4.969}

\bibitem[{Fredon {et~al.}(2021)Fredon, Radchenko, \& Cuppen}]{Fredon2021}
Fredon, A., Radchenko, A.~K., \& Cuppen, H.~M. 2021, Accounts of Chemical Research, 54, 745, \dodoi{10.1021/acs.accounts.0c00636}

\bibitem[{Frisch {et~al.}(2016)Frisch, Trucks, Schlegel, Scuseria, Robb, Cheeseman, Scalmani, Barone, Petersson, Nakatsuji, Li, Caricato, Marenich, Bloino, Janesko, Gomperts, Mennucci, Hratchian, Ortiz, Izmaylov, Sonnenberg, Williams-Young, Ding, Lipparini, Egidi, Goings, Peng, Petrone, Henderson, Ranasinghe, Zakrzewski, Gao, Rega, Zheng, Liang, Hada, Ehara, Toyota, Fukuda, Hasegawa, Ishida, Nakajima, Honda, Kitao, Nakai, Vreven, Throssell, Montgomery, Peralta, Ogliaro, Bearpark, Heyd, Brothers, Kudin, Staroverov, Keith, Kobayashi, Normand, Raghavachari, Rendell, Burant, Iyengar, Tomasi, Cossi, Millam, Klene, Adamo, Cammi, Ochterski, Martin, Morokuma, Farkas, Foresman, \& Fox}]{g16}
Frisch, M.~J., Trucks, G.~W., Schlegel, H.~B., {et~al.} 2016, Gaussian 16 {R}evision {C}.01

\bibitem[{García-Bernete {et~al.}(2021)García-Bernete, Rigopoulou, Alonso-Herrero, Pereira-Santaella, Roche, \& Kerkeni}]{GarciaBernete2021}
García-Bernete, I., Rigopoulou, D., Alonso-Herrero, A., {et~al.} 2021, Monthly Notices of the Royal Astronomical Society, 509, 4256, \dodoi{10.1093/mnras/stab3127}

\bibitem[{Gardner \& Winnewisser(1975)}]{gardner75}
Gardner, F.~F., \& Winnewisser, G. 1975, The Astrophysical Journal, 195, L127, \dodoi{10.1086/181726}

\bibitem[{{Gerakines} {et~al.}(1996){Gerakines}, {Schutte}, \& {Ehrenfreund}}]{gerakines96}
{Gerakines}, P.~A., {Schutte}, W.~A., \& {Ehrenfreund}, P. 1996, Astronomy {\&} Astrophysics, 312, 289

\bibitem[{Glavin {et~al.}(2018)Glavin, Alexander, Aponte, Dworkin, Elsila, \& Yabuta}]{glavin18}
Glavin, D.~P., Alexander, C.~M., Aponte, J.~C., {et~al.} 2018, in Primitive Meteorites and Asteroids, ed. N.~Abreu (Elsevier), 205--271, \dodoi{10.1016/B978-0-12-813325-5.00003-3}

\bibitem[{Godard {et~al.}(2012)Godard, Geballe, Dartois, \& Caro}]{Godard2012}
Godard, M., Geballe, T.~R., Dartois, E., \& Caro, G. M.~M. 2012, Astronomy {\&} Astrophysics, 537, A27, \dodoi{10.1051/0004-6361/201117197}

\bibitem[{{Goto} {et~al.}(2000){Goto}, {Maihara}, {Terada}, {Kaito}, {Kimura}, \& {Wada}}]{Goto2000}
{Goto}, M., {Maihara}, T., {Terada}, H., {et~al.} 2000, \aaps, 141, 149, \dodoi{10.1051/aas:2000113}

\bibitem[{{Goto} {et~al.}(2003){Goto}, {Gaessler}, {Hayano}, {Iye}, {Kamata}, {Kanzawa}, {Kobayashi}, {Minowa}, {Saint-Jacques}, {Takami}, {Takato}, \& {Terada}}]{goto03}
{Goto}, M., {Gaessler}, W., {Hayano}, Y., {et~al.} 2003, \apj, 589, 419, \dodoi{10.1086/368018}

\bibitem[{{Goto} {et~al.}(2007){Goto}, {Kwok}, {Takami}, {Hayashi}, {Gaessler}, {Hayano}, {Iye}, {Kamata}, {Kanzawa}, {Kobayashi}, {Minowa}, {Nedachi}, {Oya}, {Pyo}, {Saint-Jacques}, {Takato}, {Terada}, \& {Henning}}]{Goto2007}
{Goto}, M., {Kwok}, S., {Takami}, H., {et~al.} 2007, \apj, 662, 389, \dodoi{10.1086/511126}

\bibitem[{Grimme {et~al.}(2011)Grimme, Ehrlich, \& Goerigk}]{grimme11}
Grimme, S., Ehrlich, S., \& Goerigk, L. 2011, Journal of Computational Chemistry, 32, 1456, \dodoi{10.1002/jcc.21759}

\bibitem[{Gross(2017)}]{gross17}
Gross, J.~H. 2017, Fragmentation of Organic Ions and Interpretation of EI Mass Spectra (Cham: Springer International Publishing), 325--437, \dodoi{10.1007/978-3-319-54398-7_6}

\bibitem[{Günay {et~al.}(2020)Günay, Burton, Afşar, \& Schmidt}]{gunay20}
Günay, B., Burton, M.~G., Afşar, M., \& Schmidt, T.~W. 2020, Monthly Notices of the Royal Astronomical Society, 493, 1109, \dodoi{10.1093/mnras/staa288}

\bibitem[{Hansen {et~al.}(2022)Hansen, Peeters, Cami, \& Schmidt}]{hansen22}
Hansen, C.~S., Peeters, E., Cami, J., \& Schmidt, T.~W. 2022, Communications Chemistry, 5, \dodoi{10.1038/s42004-022-00714-3}

\bibitem[{Hollis {et~al.}(2004)Hollis, Jewell, Lovas, Remijan, \& M{\o}llendal}]{hollis04}
Hollis, J.~M., Jewell, P.~R., Lovas, F.~J., Remijan, A., \& M{\o}llendal, H. 2004, The Astrophysical Journal, 610, L21, \dodoi{10.1086/423200}

\bibitem[{Hoogerbrugge {et~al.}(1989)Hoogerbrugge, Bobeldijk, \& Los}]{hoogerbrugge89}
Hoogerbrugge, R., Bobeldijk, M., \& Los, J. 1989, The Journal of Physical Chemistry, 93, 5444, \dodoi{10.1021/j100351a026}

\bibitem[{{Jenniskens} {et~al.}(1993){Jenniskens}, {Baratta}, {Kouchi}, {de Groot}, {Greenberg}, \& {Strazzulla}}]{jenniskens93}
{Jenniskens}, P., {Baratta}, G.~A., {Kouchi}, A., {et~al.} 1993, \aap, 273, 583

\bibitem[{Jensen {et~al.}(2022)Jensen, Shannon, Peeters, Sloan, \& Stock}]{Jensen2022}
Jensen, P.~A., Shannon, M.~J., Peeters, E., Sloan, G.~C., \& Stock, D.~J. 2022, A\&A, 665, A153, \dodoi{10.1051/0004-6361/202141511}

\bibitem[{{Joblin} {et~al.}(1996){Joblin}, {Tielens}, {Allamandola}, \& {Geballe}}]{Joblin1996}
{Joblin}, C., {Tielens}, A.~G.~G.~M., {Allamandola}, L.~J., \& {Geballe}, T.~R. 1996, \apj, 458, 610, \dodoi{10.1086/176843}

\bibitem[{{Jones}(2012)}]{Jones2012_c}
{Jones}, A.~P. 2012, Astronomy {\&} Astrophysics, 542, A98, \dodoi{10.1051/0004-6361/201118483}

\bibitem[{Jones(2012{\natexlab{a}})}]{jones12}
Jones, A.~P. 2012{\natexlab{a}}, Astronomy {\&} Astrophysics, 540, A2, \dodoi{10.1051/0004-6361/201117624}

\bibitem[{Jones(2012{\natexlab{b}})}]{jones12_corr}
---. 2012{\natexlab{b}}, Astronomy {\&} Astrophysics, 545, C2, \dodoi{10.1051/0004-6361/201117624e}

\bibitem[{Jones(2016)}]{Jones2016}
---. 2016, Royal Society Open Science, 3, 160224, \dodoi{10.1098/rsos.160224}

\bibitem[{{Jones} {et~al.}(1990){Jones}, {Duley}, \& {Williams}}]{Jones1990}
{Jones}, A.~P., {Duley}, W.~W., \& {Williams}, D.~A. 1990, \qjras, 31, 567

\bibitem[{{Jones} {et~al.}(2013){Jones}, {Fanciullo}, {K{\"o}hler}, {Verstraete}, {Guillet}, {Bocchio}, \& {Ysard}}]{jones13}
{Jones}, A.~P., {Fanciullo}, L., {K{\"o}hler}, M., {et~al.} 2013, Astronomy and astrophysics, 558, A62, \dodoi{10.1051/0004-6361/201321686}

\bibitem[{{Jones} \& {Habart}(2015)}]{Jones2015}
{Jones}, A.~P., \& {Habart}, E. 2015, \aap, 581, A92, \dodoi{10.1051/0004-6361/201526487}

\bibitem[{{Jones} {et~al.}(2017){Jones}, {K{\"o}hler}, {Ysard}, {Bocchio}, \& {Verstraete}}]{jones17}
{Jones}, A.~P., {K{\"o}hler}, M., {Ysard}, N., {Bocchio}, M., \& {Verstraete}, L. 2017, \aap, 602, A46, \dodoi{10.1051/0004-6361/201630225}

\bibitem[{{Jones} \& {Ysard}(2022)}]{Jones2022}
{Jones}, A.~P., \& {Ysard}, N. 2022, \aap, 657, A128, \dodoi{10.1051/0004-6361/202141793}

\bibitem[{Jones {et~al.}(2011)Jones, Zhang, Kaiser, Jamal, Mebel, Cordiner, \& Charnley}]{jones_kaiser11}
Jones, B.~M., Zhang, F., Kaiser, R.~I., {et~al.} 2011, Proceedings of the National Academy of Sciences, 108, 452, \dodoi{10.1073/pnas.1012468108}

\bibitem[{Jordanov {et~al.}(2003)Jordanov, Tsankov, \& Korte}]{jordanov03}
Jordanov, B., Tsankov, D., \& Korte, E. 2003, Journal of Molecular Structure, 651-653, 101, \dodoi{https://doi.org/10.1016/S0022-2860(02)00632-4}

\bibitem[{Keller {et~al.}(2006)Keller, Bajt, Baratta, Borg, Bradley, Brownlee, Busemann, Brucato, Burchell, Colangeli, d'Hendecourt, Djouadi, Ferrini, Flynn, Franchi, Fries, Grady, Graham, Grossemy, Kearsley, Matrajt, Nakamura-Messenger, Mennella, Nittler, Palumbo, Stadermann, Tsou, Rotundi, Sandford, Snead, Steele, Wooden, \& Zolensky}]{keller06}
Keller, L.~P., Bajt, S., Baratta, G.~A., {et~al.} 2006, Science, 314, 1728, \dodoi{10.1126/science.1135796}

\bibitem[{Kwok \& Zhang(2011)}]{kwok11}
Kwok, S., \& Zhang, Y. 2011, Nature, 479, 80, \dodoi{10.1038/nature10542}

\bibitem[{{Kwok} \& {Zhang}(2013)}]{kwok13}
{Kwok}, S., \& {Zhang}, Y. 2013, \apj, 771, 5, \dodoi{10.1088/0004-637X/771/1/5}

\bibitem[{Lee {et~al.}(1988)Lee, Yang, \& Parr}]{lee88}
Lee, C., Yang, W., \& Parr, R.~G. 1988, Physical Review B, 37, 785, \dodoi{10.1103/physrevb.37.785}

\bibitem[{Lee {et~al.}(2021)Lee, Loomis, Burkhardt, Cooke, Xue, Siebert, Shingledecker, Remijan, Charnley, McCarthy, \& McGuire}]{lee21}
Lee, K. L.~K., Loomis, R.~A., Burkhardt, A.~M., {et~al.} 2021, The Astrophysical Journal, 908, L11, \dodoi{10.3847/2041-8213/abdbb9}

\bibitem[{{Leger} \& {Puget}(1984)}]{leger84}
{Leger}, A., \& {Puget}, J. 1984, A\&A, 137, L5

\bibitem[{Lewis {et~al.}(2011)Lewis, Jel{\'{\i}}nek, Ortega, Demkov, Trabada, Haycock, Wang, Adams, Tomfohr, Abad, Wang, \& Drabold}]{lewis11}
Lewis, J.~P., Jel{\'{\i}}nek, P., Ortega, J., {et~al.} 2011, physica status solidi (b), 248, 1989, \dodoi{10.1002/pssb.201147259}

\bibitem[{Li(2020)}]{Li2020}
Li, A. 2020, Nature Astronomy, 4, 339, \dodoi{10.1038/s41550-020-1051-1}

\bibitem[{Lin {et~al.}(2013)Lin, Talbi, Roueff, Herbst, Wehres, Cole, Yang, Snow, \& Bierbaum}]{lin13}
Lin, Z., Talbi, D., Roueff, E., {et~al.} 2013, The Astrophysical Journal, 765, 80, \dodoi{10.1088/0004-637x/765/2/80}

\bibitem[{Lipsky(1981)}]{lipsky81}
Lipsky, S. 1981, Journal of Chemical Education, 58, 93, \dodoi{10.1021/ed058p93}

\bibitem[{{Loeffler} {et~al.}(2005){Loeffler}, {Baratta}, {Palumbo}, {Strazzulla}, \& {Baragiola}}]{loeffler05}
{Loeffler}, M.~J., {Baratta}, G.~A., {Palumbo}, M.~E., {Strazzulla}, G., \& {Baragiola}, R.~A. 2005, \aap, 435, 587, \dodoi{10.1051/0004-6361:20042256}

\bibitem[{Loison \& Bergeat(2009)}]{loison09}
Loison, J.-C., \& Bergeat, A. 2009, Physical Chemistry Chemical Physics, 11, 655, \dodoi{10.1039/B812810C}

\bibitem[{Los {et~al.}(1991)Los, Kornig, Kistemaker, \& Beijersbergen}]{los91}
Los, J., Kornig, S., Kistemaker, P.~G., \& Beijersbergen, J. H.~M. 1991, The Journal of Physical Chemistry, 95, 2143, \dodoi{10.1021/j100159a014}

\bibitem[{Mao {et~al.}(2019)Mao, Kroll, \& Schug}]{mao19}
Mao, J.~X., Kroll, P., \& Schug, K.~A. 2019, Structural Chemistry, 30, 2217, \dodoi{10.1007/s11224-019-01412-y}

\bibitem[{Marcelino {et~al.}(2007)Marcelino, Cernicharo, Ag{\'{u}}ndez, Roueff, Gerin, Mart{\'{\i}}n-Pintado, Mauersberger, \& Thum}]{marcelino07}
Marcelino, N., Cernicharo, J., Ag{\'{u}}ndez, M., {et~al.} 2007, The Astrophysical Journal, 665, L127, \dodoi{10.1086/521398}

\bibitem[{Marciniak {et~al.}(2021)Marciniak, Joblin, Mulas, Mundlapati, \& Bonnamy}]{marciniak21}
Marciniak, A., Joblin, C., Mulas, G., Mundlapati, V.~R., \& Bonnamy, A. 2021, Astronomy {\&} Astrophysics, 652, A42, \dodoi{10.1051/0004-6361/202140737}

\bibitem[{Mart{\'{\i}}n-Dom{\'{e}}nech {et~al.}(2015)Mart{\'{\i}}n-Dom{\'{e}}nech, Manzano-Santamar{\'{\i}}a, Caro, Cruz-D{\'{\i}}az, Chen, Herrero, \& Tanarro}]{MartinDomenech15}
Mart{\'{\i}}n-Dom{\'{e}}nech, R., Manzano-Santamar{\'{\i}}a, J., Caro, G. M.~M., {et~al.} 2015, Astronomy {\&} Astrophysics, 584, A14, \dodoi{10.1051/0004-6361/201526003}

\bibitem[{Mart{\'{\i}}nez {et~al.}(2020)Mart{\'{\i}}nez, Santoro, Merino, Accolla, Lauwaet, Sobrado, Sabbah, Pelaez, Herrero, Tanarro, Ag{\'{u}}ndez, Mart{\'{\i}}n-Jimenez, Otero, Ellis, Joblin, Cernicharo, \& Mart{\'{\i}}n-Gago}]{martinez20}
Mart{\'{\i}}nez, L., Santoro, G., Merino, P., {et~al.} 2020, Nature Astronomy, 4, 97, \dodoi{10.1038/s41550-019-0899-4}

\bibitem[{{Mathis} {et~al.}(1983){Mathis}, {Mezger}, \& {Panagia}}]{mathis83}
{Mathis}, J.~S., {Mezger}, P.~G., \& {Panagia}, N. 1983, \aap, 128, 212

\bibitem[{{Matrajt} {et~al.}(2005){Matrajt}, {Mu{\~n}oz Caro}, {Dartois}, {D'Hendecourt}, {Deboffle}, \& {Borg}}]{matrajt05}
{Matrajt}, G., {Mu{\~n}oz Caro}, G.~M., {Dartois}, E., {et~al.} 2005, \aap, 433, 979, \dodoi{10.1051/0004-6361:20041605}

\bibitem[{McCarthy {et~al.}(2021)McCarthy, Lee, Loomis, Burkhardt, Shingledecker, Charnley, Cordiner, Herbst, Kalenskii, Willis, Xue, Remijan, \& McGuire}]{mccarthy21_b}
McCarthy, M.~C., Lee, K. L.~K., Loomis, R.~A., {et~al.} 2021, Nature Astronomy, 5, 176, \dodoi{10.1038/s41550-020-01213-y}

\bibitem[{McGuire(2022)}]{mcguire22}
McGuire, B.~A. 2022, The Astrophysical Journal Supplement Series, 259, 30, \dodoi{10.3847/1538-4365/ac2a48}

\bibitem[{McGuire {et~al.}(2021)McGuire, Loomis, Burkhardt, Lee, Shingledecker, Charnley, Cooke, Cordiner, Herbst, Kalenskii, Siebert, Willis, Xue, Remijan, \& McCarthy}]{mcguire21}
McGuire, B.~A., Loomis, R.~A., Burkhardt, A.~M., {et~al.} 2021, Science, 371, 1265, \dodoi{10.1126/science.abb7535}

\bibitem[{Meinert {et~al.}(2016)Meinert, Myrgorodska, de~Marcellus, Buhse, Nahon, Hoffmann, d’Hendecourt, \& Meierhenrich}]{Meinert16}
Meinert, C., Myrgorodska, I., de~Marcellus, P., {et~al.} 2016, Science, 352, 208, \dodoi{10.1126/science.aad8137}

\bibitem[{Mennella {et~al.}(2001)Mennella, Caro, Ruiterkamp, Schutte, Greenberg, Brucato, \& Colangeli}]{mennella01}
Mennella, V., Caro, G. M.~M., Ruiterkamp, R., {et~al.} 2001, Astronomy {\&} Astrophysics, 367, 355, \dodoi{10.1051/0004-6361:20000340}

\bibitem[{Molpeceres \& Rivilla(2022)}]{molpeceres22}
Molpeceres, G., \& Rivilla, V.~M. 2022, Astronomy {\&} Astrophysics, 665, A27, \dodoi{10.1051/0004-6361/202243892}

\bibitem[{Monfredini {et~al.}(2019)Monfredini, Quitián-Lara, Fantuzzi, Wolff, Mendoza, Lago, Sales, Pastoriza, \& Boechat-Roberty}]{Monfredini2019}
Monfredini, T., Quitián-Lara, H.~M., Fantuzzi, F., {et~al.} 2019, Monthly Notices of the Royal Astronomical Society, 488, 451, \dodoi{10.1093/mnras/stz1021}

\bibitem[{Morales {et~al.}(2011)Morales, Bennett, Picard, Canosa, Sims, Sun, Chen, Chang, Kislov, Mebel, Gu, Zhang, Maksyutenko, \& Kaiser}]{morales11}
Morales, S.~B., Bennett, C.~J., Picard, S. D.~L., {et~al.} 2011, The Astrophysical Journal, 742, 26, \dodoi{10.1088/0004-637x/742/1/26}

\bibitem[{Morisawa {et~al.}(2012)Morisawa, Tachibana, Ehara, \& Ozaki}]{morisawa12}
Morisawa, Y., Tachibana, S., Ehara, M., \& Ozaki, Y. 2012, The Journal of Physical Chemistry A, 116, 11957, \dodoi{10.1021/jp307634m}

\bibitem[{{Mu{\~n}oz Caro} {et~al.}(2001){Mu{\~n}oz Caro}, {Ruiterkamp}, {Schutte}, {Greenberg}, \& {Mennella}}]{caro01}
{Mu{\~n}oz Caro}, G.~M., {Ruiterkamp}, R., {Schutte}, W.~A., {Greenberg}, J.~M., \& {Mennella}, V. 2001, \aap, 367, 347, \dodoi{10.1051/0004-6361:20000341}

\bibitem[{Murga {et~al.}(2023)Murga, Vasyunin, \& Kirsanova}]{murga23}
Murga, M.~S., Vasyunin, A.~I., \& Kirsanova, M.~S. 2023, Monthly Notices of the Royal Astronomical Society, 519, 2466, \dodoi{10.1093/mnras/stac3656}

\bibitem[{Peeters {et~al.}(2021)Peeters, Mackie, Candian, \& Tielens}]{peeters21}
Peeters, E., Mackie, C., Candian, A., \& Tielens, A. G. G.~M. 2021, Accounts of Chemical Research, 54, 1921, \dodoi{10.1021/acs.accounts.0c00747}

\bibitem[{Pendleton \& Allamandola(2002)}]{pendleton02}
Pendleton, Y.~J., \& Allamandola, L.~J. 2002, The Astrophysical Journal Supplement Series, 138, 75, \dodoi{10.1086/322999}

\bibitem[{Pety {et~al.}(2005)Pety, Teyssier, Foss{\'{e}}, Gerin, Roueff, Abergel, Habart, \& Cernicharo}]{pety05}
Pety, J., Teyssier, D., Foss{\'{e}}, D., {et~al.} 2005, Astronomy {\&} Astrophysics, 435, 885, \dodoi{10.1051/0004-6361:20041170}

\bibitem[{{Pinho} \& {Duley}(1995)}]{Pinho1995}
{Pinho}, G.~P., \& {Duley}, W.~W. 1995, \apjl, 442, L41, \dodoi{10.1086/187811}

\bibitem[{{Pino} {et~al.}(2008){Pino}, {Dartois}, {Cao}, {Carpentier}, {Chamaill{\'e}}, {Vasquez}, {Jones}, {D'Hendecourt}, \& {Br{\'e}chignac}}]{pino08}
{Pino}, T., {Dartois}, E., {Cao}, A.~T., {et~al.} 2008, \aap, 490, 665, \dodoi{10.1051/0004-6361:200809927}

\bibitem[{{Puget} \& {Leger}(1989)}]{puget89}
{Puget}, J., \& {Leger}, A. 1989, Annual Review of Astronomy and Astrophysics, 27, 161

\bibitem[{{Raponi} {et~al.}(2020){Raponi}, {Ciarniello}, {Capaccioni}, {Mennella}, {Filacchione}, {Vinogradoff}, {Poch}, {Beck}, {Quirico}, {De Sanctis}, {Moroz}, {Kappel}, {Erard}, {Bockel{\'e}e-Morvan}, {Longobardo}, {Tosi}, {Palomba}, {Combe}, {Rousseau}, {Arnold}, {Carlson}, {Pommerol}, {Pilorget}, {Fornasier}, {Bellucci}, {Barucci}, {Mancarella}, {Formisano}, {Rinaldi}, {Istiqomah}, \& {Leyrat}}]{raponi20}
{Raponi}, A., {Ciarniello}, M., {Capaccioni}, F., {et~al.} 2020, Nature Astronomy, 4, 500, \dodoi{10.1038/s41550-019-0992-8}

\bibitem[{Rivilla {et~al.}(2022)Rivilla, Colzi, Jim{\'{e}}nez-Serra, Mart{\'{\i}}n-Pintado, Meg{\'{\i}}as, Melosso, Bizzocchi, L{\'{o}}pez-Gallifa, Mart{\'{\i}}nez-Henares, Massalkhi, Tercero, de~Vicente, Guillemin, de~la Concepci{\'{o}}n, Rico-Villas, Zeng, Mart{\'{\i}}n, Requena-Torres, Tonolo, Alessandrini, Dore, Barone, \& Puzzarini}]{rivilla22}
Rivilla, V.~M., Colzi, L., Jim{\'{e}}nez-Serra, I., {et~al.} 2022, The Astrophysical Journal Letters, 929, L11, \dodoi{10.3847/2041-8213/ac6186}

\bibitem[{Santoro {et~al.}(2020{\natexlab{a}})Santoro, Sobrado, Tajuelo-Castilla, Accolla, Mart{\'{\i}}nez, Azpeitia, Lauwaet, Cernicharo, Ellis, \& Mart{\'{\i}}n-Gago}]{santoro20rev}
Santoro, G., Sobrado, J.~M., Tajuelo-Castilla, G., {et~al.} 2020{\natexlab{a}}, Review of Scientific Instruments, 91, 124101, \dodoi{10.1063/5.0027920}

\bibitem[{Santoro {et~al.}(2020{\natexlab{b}})Santoro, Mart{\'{\i}}nez, Lauwaet, Accolla, Tajuelo-Castilla, Merino, Sobrado, Pel{\'{a}}ez, Herrero, Tanarro, Mayoral, Ag{\'{u}}ndez, Sabbah, Joblin, Cernicharo, \& Mart{\'{\i}}n-Gago}]{santoro20}
Santoro, G., Mart{\'{\i}}nez, L., Lauwaet, K., {et~al.} 2020{\natexlab{b}}, The Astrophysical Journal, 895, 97, \dodoi{10.3847/1538-4357/ab9086}

\bibitem[{{Schuhmann, M.} {et~al.}(2019){Schuhmann, M.}, {Altwegg, K.}, {Balsiger, H.}, {Berthelier, J.-J.}, {De Keyser, J.}, {Fiethe, B.}, {Fuselier, S. A.}, {Gasc, S.}, {Gombosi, T. I.}, {H\"anni, N.}, {Rubin, M.}, {Tzou, C.-Y.}, \& {Wampfler, S. F.}}]{schuhmann19}
{Schuhmann, M.}, {Altwegg, K.}, {Balsiger, H.}, {et~al.} 2019, A\&A, 630, A31, \dodoi{10.1051/0004-6361/201834666}

\bibitem[{Smith(1984)}]{Smith1984}
Smith, F.~W. 1984, Journal of Applied Physics, 55, 764, \dodoi{10.1063/1.333135}

\bibitem[{Smith {et~al.}(2006)Smith, Sage, Donahue, Herbst, \& Quan}]{smith06}
Smith, I. W.~M., Sage, A.~M., Donahue, N.~M., Herbst, E., \& Quan, D. 2006, Faraday Discussions, 133, 137, \dodoi{10.1039/b600721j}

\bibitem[{Snyder {et~al.}(1978)Snyder, Hsu, \& Krimm}]{snyder78}
Snyder, R., Hsu, S., \& Krimm, S. 1978, Spectrochimica Acta Part A: Molecular Spectroscopy, 34, 395 , \dodoi{https://doi.org/10.1016/0584-8539(78)80167-6}

\bibitem[{Snyder \& Schachtschneider(1963)}]{snyder63}
Snyder, R., \& Schachtschneider, J. 1963, Spectrochimica Acta, 19, 85 , \dodoi{https://doi.org/10.1016/0371-1951(63)80095-8}

\bibitem[{Sobrado {et~al.}(2023)Sobrado, Santoro, Mart{\'i}nez, Merino, Joblin, Cernicharo, \& Gago}]{sobrado23}
Sobrado, J., Santoro, G., Mart{\'i}nez, L., {et~al.} 2023, in European Conference on Laboratory Astrophysics ECLA2020, ed. V.~Mennella \& C.~Joblin (Cham: Springer International Publishing), 101--110

\bibitem[{Socrates(2004)}]{socrates}
Socrates, G. 2004, Infrared and Raman Characteristic Group Frequencies: Tables and Charts, 3rd edn.

\bibitem[{Steenvoorden {et~al.}(1991)Steenvoorden, Kistemaker, {De Vries}, Michalak, \& Nibbering}]{steenvoorden91}
Steenvoorden, R., Kistemaker, P., {De Vries}, A., Michalak, L., \& Nibbering, N. 1991, International Journal of Mass Spectrometry and Ion Processes, 107, 475, \dodoi{https://doi.org/10.1016/0168-1176(91)80042-L}

\bibitem[{Tielens(2008)}]{tielens08}
Tielens, A. 2008, Annual Review of Astronomy and Astrophysics, 46, 289, \dodoi{10.1146/annurev.astro.46.060407.145211}

\bibitem[{Tielens(2005{\natexlab{a}})}]{tielens05_book}
Tielens, A. G. G.~M. 2005{\natexlab{a}}, The Physics and Chemistry of the Interstellar Medium (Cambridge University Press), \dodoi{10.1017/CBO9780511819056}

\bibitem[{Tielens(2005{\natexlab{b}})}]{tielens05_ch}
---. 2005{\natexlab{b}}, Interstellar polycyclic aromatic hydrocarbon molecules (Cambridge University Press), 173–227, \dodoi{10.1017/CBO9780511819056.007}

\bibitem[{{Tielens}(2013)}]{tielens13}
{Tielens}, A.~G.~G.~M. 2013, Reviews of Modern Physics, 85, 1021, \dodoi{10.1103/RevModPhys.85.1021}

\bibitem[{Yabuta {et~al.}(2023)Yabuta, Cody, Engrand, Kebukawa, Gregorio, Bonal, Remusat, Stroud, Quirico, Nittler, Hashiguchi, Komatsu, Okumura, Mathurin, Dartois, Duprat, Takahashi, Takeichi, Kilcoyne, Yamashita, Dazzi, Deniset-Besseau, Sandford, Martins, Tamenori, Ohigashi, Suga, Wakabayashi, Verdier-Paoletti, Mostefaoui, Montagnac, Barosch, Kamide, Shigenaka, Bejach, Matsumoto, Enokido, Noguchi, Yurimoto, Nakamura, Okazaki, Naraoka, Sakamoto, Connolly, Lauretta, Abe, Okada, Yada, Nishimura, Yogata, Nakato, Yoshitake, Iwamae, Furuya, Hatakeda, Miyazaki, Soejima, Hitomi, Kumagai, Usui, Hayashi, Yamamoto, Fukai, Sugita, Kitazato, Hirata, Honda, Morota, Tatsumi, Sakatani, Namiki, Matsumoto, Noguchi, Wada, Senshu, Ogawa, Yokota, Ishihara, Shimaki, Yamada, Honda, Michikami, Matsuoka, Hirata, Arakawa, Okamoto, Ishiguro, Jaumann, Bibring, Grott, Schröder, Otto, Pilorget, Schmitz, Biele, Ho, Moussi-Soffys, Miura, Noda, Yamada, Yoshihara, Kawahara, Ikeda, Yamamoto, Shirai, Kikuchi, Ogawa, Takeuchi, Ono, Mimasu,
  Yoshikawa, Takei, Fujii, ichi Iijima, Nakazawa, Hosoda, Iwata, Hayakawa, Sawada, Yano, Tsukizaki, Ozaki, Terui, Tanaka, Fujimoto, Yoshikawa, Saiki, Tachibana, ichiro Watanabe, \& Tsuda}]{yabuta23}
Yabuta, H., Cody, G.~D., Engrand, C., {et~al.} 2023, Science, 379, eabn9057, \dodoi{10.1126/science.abn9057}

\bibitem[{Yang {et~al.}(2013)Yang, Glaser, Li, \& Zhong}]{Yang2013}
Yang, X.~J., Glaser, R., Li, A., \& Zhong, J.~X. 2013, The Astrophysical Journal, 776, 110, \dodoi{10.1088/0004-637x/776/2/110}

\bibitem[{Yang \& Li(2023)}]{Yang2023}
Yang, X.~J., \& Li, A. 2023, The Astrophysical Journal Supplement Series, 268, 50, \dodoi{10.3847/1538-4365/acebe6}

\bibitem[{Ysard {et~al.}(2015)Ysard, Köhler, Jones, Miville-Deschênes, Abergel, \& Fanciullo}]{Ysard2015}
Ysard, N., Köhler, M., Jones, A., {et~al.} 2015, Astronomy {\&} Astrophysics, 577, A110, \dodoi{10.1051/0004-6361/201425523}

\bibitem[{Öberg(2016)}]{oberg16}
Öberg, K.~I. 2016, Chemical Reviews, 116, 9631, \dodoi{10.1021/acs.chemrev.5b00694}

\end{thebibliography}
\bibliographystyle{aasjournal}

\end{document}